\documentclass[prd,nofootinbib,floats,aps,twocolumn,tightenlines,superscriptaddress]{revtex4-2}
\usepackage{natbib}
\usepackage{graphicx}
\usepackage{mathtools}
\usepackage{hyperref}
\hypersetup{
    colorlinks=true,
    linkcolor=blue,
    filecolor=magenta,      
    urlcolor=cyan,
}
\usepackage{commath}
\usepackage{bm}
\usepackage{amsfonts,amsmath,amssymb,amsthm}
\usepackage[usenames,dvipsnames]{xcolor}
\usepackage[utf8]{inputenc}
\usepackage[T1]{fontenc}
\usepackage{subfigure}
\usepackage{dsfont}
\usepackage{enumitem}
\usepackage{qcircuit}
\usepackage{color}
\usepackage{tikz}
\usepackage{physics}
\usepackage{verbatim}
\usepackage[normalem]{ulem}
\newcommand{\ii}{\mathrm{i}}

\renewcommand*\d[2][]{%
	\mathrm{d}%
	\ifx\relax#1\relax\else
	\rule{-0.02em}{1.5ex}^{#1}\rule{0.08em}{0ex}\!
	\fi
	#2\,
}

\newcommand{\bigket}[1]{\big|#1\big\rangle}
\newcommand{\bigbra}[1]{\big\langle#1\big|}
\newcommand{\Bigket}[1]{\Big|#1\Big\rangle}
\newcommand{\Bigbra}[1]{\Big\langle#1\Big|}
\newcommand{\rhoh}{\hat \rho}
\newcommand{\phih}{\hat \phi}

\renewcommand{\dd}{\mathrm{d}}
\newcommand{\tcr}{\textcolor{red}}

\newcommand{\ta}{\textsc{a}}
\newcommand{\tb}{\textsc{b}}
\newcommand{\tab}{\textsc{ab}}
\newcommand{\tf}{\textsc{f}}
\newcommand{\ti}{\textsc{i}}

\begin{document}

\title{Sabotaging the harvesting of correlations from quantum fields}
	
\author{Abhisek Sahu}
 \email{abhi@phas.ubc.ca}
\affiliation{Department of Physics and Astronomy, University of British Columbia, 6224 Agricultural Road, Vancouver, B.C. V6T 1Z1, Canada}

\author{Irene Melgarejo-Lermas}
\email{i2melgar@uwaterloo.ca}
\affiliation{Department of Applied Mathematics, University of Waterloo, Waterloo, Ontario, N2L 3G1, Canada}
\affiliation{Institute for Quantum Computing, University of Waterloo, Waterloo, Ontario, N2L 3G1, Canada}

\author{Eduardo Mart\'{i}n-Mart\'{i}nez}
 \email{emartinmartinez@uwaterloo.ca}
\affiliation{Department of Applied Mathematics, University of Waterloo, Waterloo, Ontario, N2L 3G1, Canada}
\affiliation{Institute for Quantum Computing, University of Waterloo, Waterloo, Ontario, N2L 3G1, Canada}
\affiliation{Perimeter Institute for Theoretical Physics, Waterloo, Ontario N2L 2Y5, Canada}

\begin{abstract}

We study the non-perturbative harvesting of classical and quantum correlations between two parties coupled to a quantum field. First, we consider a scenario with an arbitrary number of two-level systems that couple to a quantum field locally in time. Then, we study the impact of the presence of additional detectors (interlopers) on the ability for two target detectors (Alice and Bob) to acquire  correlations through their interaction with the field.  
We analyze the harvesting of different correlation measures in this non-perturbative regime and we demonstrate that even a single interloper can completely sabotage all correlation harvesting between Alice and Bob by acting on the causal past of one of them. Specifically, we show that the interloper is able to interact with the field so that the field itself `floods' one of the parties with entropy. This prevents Alice and Bob from acquiring any correlations. Furthermore, we show that this kind of attack cannot be defended against. 


\end{abstract}
	
\maketitle

\section{Introduction}


It is a well-known fact that quantum fields contain correlations between time-like and space-like separated regions. For example, Summers and Werner showed that the vacuum state of a free quantum field displays spacelike entanglement~\cite{Werner1,Werner2}.
The study of the entanglement structure of quantum field states is at the centre of fundamental questions such as the black hole information loss problem~\cite{Preskill}, and is  an interesting focus of study in itself. 
More importantly, the presence of this entanglement has been utilized to perform relativistic quantum informational  tasks such as \textit{quantum energy teleportation} \cite{Hotta_2008} and \textit{entanglement harvesting}.

Indeed, the word harvesting is used because the entanglement contained in the field can be extracted to particle detectors (for example, two hydrogen atoms can harvest entanglement and other correlations from the vacuum state of the electromagnetic field~\cite{hydrogenpozas}). The extraction of those correlations between spacelike separated detectors has become known generically under the name of entanglement (or correlation)  harvesting. Entanglement harvesting was originally explored by Valentini~\cite{VALENTINI1991321} and Reznik~\cite{Reznik2003-REZEFT}, and since then the analysis have been broadened to the general extraction of correlations~\cite{corrvac} and quantum discord~\cite{brown2013,Borrelli_2012} as well. The idea behind the harvesting of correlations is to consider two initially uncorrelated quantum systems (such as qubits, atoms or quantum harmonic oscillators) that interact with a quantum field for a certain amount of time.
The resultant post-interaction reduced state of the two probes displays probe-probe correlations  even when these were in spacelike separation throughout the whole interaction process. This is possible because the probes acquire or \textit{harvest} the correlations present in the quantum field state. 

Correlation harvesting protocols have been studied extensively in a plethora of distinct scenarios. In flat spacetime it has been shown that entanglement can be harvested from coherent field states \cite{nonper}, squeezed coherent states~\cite{harvcorrel}, thermal states~\cite{brown2013, harvcorrel} and also from electromagnetic vacuum using fully featured hydrogen like atoms~\cite{hydrogenpozas}.
The protocol has also been found to be sensitive to the localization of the detectors~\cite{corrvac} and their trajectories~\cite{Salton_2015}, the boundary conditions of the field \cite{seismology}, and the nature of the detector-field couplings \cite{allison2017}, as well as the geometry \cite{steegMeni,gallockyoshimura2021harvesting} and topology~\cite{vacuumEntangleEdu2016} of the background spacetime.

Many of the analyses of entanglement harvesting have been done within perturbation theory, but different non-perturbative techniques do exist (see, among others, \cite{nonper,Brown_2013,Bruschi_2013,vriend2020unruh,masahiroEnergyExtraction,Masajo}). In particular, the techniques in~\cite{nonper}  have been used to derive general no-go theorems exploring the limits of entanglement harvesting~\cite{nogo}.
Several practical applications of the protocol have been explored leading to proposals going from quantum resource production~\cite{farming}, precise detection of vibrations~\cite{seismology}, rangefinding~\cite{Salton_2015} to communication~\cite{Kojinada}.

In this paper we address the question of how robust and secure correlation harvesting protocols can be. In particular if detectors can get coupled to a field and extract correlations from it, could similar malicious detectors also couple to the field and sabotage their harvesting protocol?
To answer this question adequately, one must consider a situation where a large number of detectors interact with the field in ways more involved than those reported in previous literature known to the authors.

As a first step in answering these questions  we study the non-perturbative harvesting of correlations in the presence of an arbitrarily large number of detectors and unveil a scenario in which it is possible for a third party to easily and completely sabotage all correlations between the target detectors. 
More precisely, we study the coupling of $N$ Unruh-DeWitt detectors to a coherent field state of a scalar theory and non-perturbatively calculate the post interaction reduced state of the detectors. 
We subsequently show that even a single additional detector (henceforth referred to as \textit{interloper}) can cancel all correlations between our target detectors by acting on the causal past of one of the detectors while evading any attempts to counter their sabotage.

This manuscript is organized as follows: in Section \ref{setupA} we review the Unruh-Dewitt detector model followed by a non-perturbative calculation of the time-evolution unitary and the density matrix of the degrees of freedom corresponding to $A$, $B$ and the $N$ \textit{interloper} detectors in Section \ref{setupB}, under the assumption of a delta-like time switching function.
In Section \ref{result} we present an analytical condition to be satisfied by the interloper for cancelling the harvesting of correlations.
We also formulate general rules to be followed for the condition to be satisfied and provide examples of specific instances of cancellation in 3+1 spacetime dimensions. 
In Section \ref{sectionHarvestingCorrelations}, we calculate 1)correlators of detector observables, 2)mutual information and 3)quantum discord. In Section \ref{sectionTrendsinCorrelation} we analyze the trends of different correlation measures considering 1) the coupling strength of detectors $A$ and $B$, 2)the relative position of detectors $A$ and $B$ and 3)how multiple \textit{interloper} affect the correlation harvesting between detectors $A$ and $B$ . 
In \ref{sectionTheft} we show that the interloper does not steal the correlation between the targets, but simply sabotages them without any personal benefit. 
Finally in \ref{sectionConclusion}, we present a summary of our main results and concluding remarks.

\section{Setup} 
\subsection{The Model}
\label{setupA}
Let us consider a real massless scalar field in a $(n+1)$ dimension flat space-time. We can write it in terms of plane wave solutions to the Klein-Gordon equation as
\begin{small}
\begin{equation}
\label{field}
    \hat{\phi}(t,\bm{x}) = \int \dd^n\bm{k} \frac{1}{\sqrt{2(2\pi)^n|\bm{k}|}}\Bigg[ \hat{a}_{\bm{k}}^\dagger e^{\ii(|\bm{k}|t - \bm{k}\cdot\bm{x})}+ \hat{a}_{\bm{k}} e^{-\ii(|\bm{k}|t - \bm{k}\cdot\bm{x})} \Bigg],
\end{equation}
\end{small}  
where the creation, $\hat{a}_{\bm{k}}^\dagger$, and annihilation operators, $\hat{a}_{\bm{k}}$,  obey the canonical commutation relations
\begin{align}
    [\hat{a}_{\bm{k}},\hat{a}_{\bm{k}'}] =  [\hat{a}_{\bm{k}}^\dagger,\hat{a}_{\bm{k}'}^\dagger]=  0, \quad \quad  [\hat{a}_{\bm{k}},\hat{a}_{\bm{k}'}^\dagger]= \delta^n(\bm{k}-\bm{k}').
\end{align}
We now consider $N$ particle detectors coupled linearly to the field according to the Unruh-DeWitt model \cite{2010grae}. In this model, a particle detector is treated as a first-quantized  two-level system whose coupling to the field is localized in space and time. The interaction Hamiltonian (in the interaction picture) for the $\nu$-th detector is given by
\begin{equation}\label{Ham1det}
    \hat{H}_\nu (t) = {\lambda}_\nu \chi_\nu(t) \hat{m}_\nu(t) \otimes \int \dd^n\bm{x} F_\nu (\bm{x}-\bm{x}_\nu) \hat{\phi}(t,\bm{x}),
\end{equation}
so that all detectors $\nu\in\{1,\dots,N\}$ are co-moving with the quantization frame $(t,\bm x)$ and their centres of mass are localised at positions $\bm{x}_\nu$.  The spatial profile of the detectors are given by the real-valued distributions $F_\nu(\bm{x})$, henceforth referred to as the \textit{smearing functions.} Here, $\hat{m}_\nu(t) := \ket{g_\nu}\!\bra{e_\nu}e^{\ii \Omega_\nu t} + \ket{e_\nu}\!\bra{g_\nu}e^{-\ii \Omega_\nu t}$ is the detector's monopole moment ($\ket{g_\nu}, \ket{e_\nu}$ denote, respectively, ground and excited states and $\Omega_\nu$ is the energy gap between them). The free Hamiltonian for detectors is therefore
\begin{equation}
\hat H^{\text{free}}_\nu=\Omega_\nu \ket{e_\nu}\!\bra{e_\nu}=\frac{\Omega_\nu}{2}(\hat\sigma_{z,\nu}+\openone).
\label{pajarraco}
\end{equation}
${\lambda}_\nu$ is each detector's coupling strength. The time dependence of the coupling is controlled by the switching functions $\chi_\nu(t)$. This model captures all the essential features of a light-matter interaction  as long as exchange of angular momentum between the field and detectors can be ignored \cite{wavepacket2013, Casimir_2014}.
The interaction Hamiltonian of all $N$ detectors with the field is therefore
\begin{equation}
    \hat{H}(t) = \sum_{\nu=1}^N {\lambda}_\nu \chi_\nu(t) \hat{\mu}_\nu(t)\otimes \int \dd^n\bm{x} F_\nu (\bm{x}-\bm{x_\nu}) \hat{\phi}(t,\bm{x}),
    \label{TotalHam}
\end{equation}
where $\hat{\mu}_\nu(t) = \openone_1\otimes\dots\openone_{\nu-1}\otimes \hat{m}_\nu(t)\otimes\dots\otimes\openone_N$. For brevity, we will naturally extend operators in the Hilbert space $\mathcal{H}_\nu \otimes \mathcal{H}_\phi$ to operators in $\otimes_\nu \mathcal{H}_\nu \otimes \mathcal{H}_\phi$ by dropping the tensor products of identity operators. 

The time evolution generated by \eqref{TotalHam} is implemented by its time-ordered exponential
\begin{equation}
\label{UnitaryDef}
    \hat{U} = \mathcal{T}\exp\left[-\ii\int_{-\infty}^\infty \!\!\!\dd t \hat{H}(t)\right],
\end{equation}
so that if the initial state of the detector-field system is given by the density operator $\hat{\rho}_0$, the final state is given by
\begin{equation}
    \hat{\rho} = \hat{U}\hat{\rho}_0\hat{U}^\dagger.
\end{equation}
The final state of  the detectors after the interaction can be obtained by tracing over the field degrees of freedom $\hat{\rho}_{\textsc{d}} = \Tr_{\hat{\phi}}[\hat{\rho}]$. 
We can also obtain the density matrix of any detector or group of detectors from the above expression by tracing over the appropriate detector degrees of freedom.

\subsection{Non-perturbative time evolution of $N$ detectors}
\label{setupB}
The problem of calculating the unitary $\hat{U}$ in \eqref{UnitaryDef} is very commonly approached perturbatively, where the expansion of the exponential is carried out till certain power of ${\lambda}_\nu$. In some special cases, a non-perturbative calculation of \eqref{UnitaryDef} is possible. For example in \cite{nonper},  a delta-switching function $ \chi_\nu(t):= \eta_\nu\delta(t-t_\nu)$ has been used to find an exact expression for $\hat{U}$ and study the harvesting of correlations from coherent field states using two detectors.
The delta-switching function captures the limit of a very strong and short interaction. $\eta_\nu$ has the dimensions of length and denotes the strength of the interaction. 
We follow the approach in \cite{nonper} to generalize to the case of N detectors interacting with a coherent field state, via single delta switching functions and study the harvesting of classical and quantum correlations (recall that although it has been shown that single delta coupling cannot harvest entanglement~\cite{nogo}, we will see that quantum discord can still be harvested with them).

In this section we present an exact expression for the density matrix of all detectors and the reduced density matrix of any two. Initially the detectors are taken to be in their respective ground states.
We consider the field initialized in an arbitrary coherent state, and the initial state of the field-detectors system to be $\hat\rho_0=\ket{\psi_0}\!\bra{\psi_0}$ with
\begin{equation}\label{initstate}
    \ket{\psi_0}= \bigotimes_{\nu=1}\ket{g_\nu} \otimes \ket{\beta_0(\bm k)}.
\end{equation} 
The coherent state $\ket{\beta_0(\bm k)}$ is characterized by a coherent amplitude distribution $\beta_0(\bm{k})$ defined as 
\begin{align}
\label{coherent field state}
   \nonumber  \ket{\beta_0(\bm{k})}& = \hat{D}_{\beta_0(\bm{k})} \ket{0} \\
       &= \exp(\int \dd^n \bm{k} \big[ \beta_0(\bm{k})\hat{a}_{\bm{k}}^\dagger - \beta_0(\bm{k})^*\hat{a}_{\bm{k}}\big ])\ket{0},
\end{align}
 where $\hat{D}_{\beta_0(\bm{k})}$ is a multimode displacement operator~\cite{nonper}. Note that the vacuum state of the field is the coherent state with distribution $\beta_0(\bm{k}) = 0$ $\forall \bm k$.
 The various properties of coherent states have been reviewed in Appendix \ref{appendix: coherent states}.
 
The initial state \eqref{initstate}  will evolve in the interaction picture as $\ket{\psi} = \hat{U}\ket{\psi_0}$.
Subsequently, we can obtain the state of the detectors by tracing over the field degrees of freedom:
\begin{equation}
\label{Tracing channel}
  \hat{\rho}_{\textsc{d}}= \Tr_{\hat\phi}[\ket{\psi}\bra{\psi}]. 
\end{equation}
Each detector couples with the field through a delta-switching function,
\begin{equation}\label{switching}
    \chi_\nu(t)=\eta_\nu\delta(t-t_\nu),
\end{equation}
where the constant $\eta_\nu$ quantifies the strength of the kick~\cite{nonper}. We can define an effective coupling strength, $\tilde\lambda_\nu \coloneqq\eta_\nu \lambda_\nu$,  that we will use for notational brevity from now on.
Also, unless explicitly stated otherwise, we will work in the convenient basis $\{\ket{1_\nu}, \ket{-1_\nu}\} $ defined by
\begin{align}
\label{new basis}
    &\ket{1_\nu} \coloneqq \frac{1}{\sqrt{2}}\big(\ket{g_\nu} + e^{\ii\Omega_\nu t_\nu}\ket{e_\nu}\big), \\
    &\ket{-1_\nu} \coloneqq \frac{1}{\sqrt{2}}\big(\ket{g_\nu} - e^{\ii\Omega_\nu t_\nu}\ket{e_\nu}\big), 
\end{align}
 that we will denote throughout as $\{\ket{s_\nu}\}$, $s_\nu=\pm1$.

Considering the switching function \eqref{switching}, the expression for $\hat{U}$ in \eqref{UnitaryDef} can be written as
\begin{equation}
\label{TimeOrderUni}
    \hat{U} = \mathcal{T} \exp[-\ii\sum_{\nu=1}^N\hat{H}_\nu],
\end{equation}
where, 
\begin{equation}\label{Hnu}
\hat{H}_\nu=\ii\ \hat{S}_3^\nu\otimes \int \dd^n \bm{k} \big[ \beta_\nu(\bm{k})\hat{a}_{\bm{k}}^\dagger - \beta_\nu(\bm{k})^*\hat{a}_{\bm{k}}\big ],
\end{equation}
$\hat S_3^\nu \coloneqq   -\ket{-1_\nu}\bra{-1_\nu}+\ket{1_\nu}\bra{1_\nu}$,  the function $\beta_\nu(\bm{k})$ is given by
\begin{equation}
\label{betaNuDef}
     \beta_\nu(\bm{k})= -\ii\tilde\lambda_\nu\frac{\Tilde{F_\nu}(-\bm{k})}{\sqrt{2|\bm{k}|}}e^{\ii(|\bm{k}|t_\nu - \bm{k}\cdot\bm{x_\nu})},
\end{equation}
and
\begin{equation}
\label{fourierTransform}
\tilde{F}_\nu(\bm{k}) \coloneqq  \frac{1}{\sqrt{(2\pi)^n}}\int \dd^n\bm{x}F(\bm{x})e^{\ii\bm{k}\cdot\bm{x}}     
\end{equation}
 is the Fourier transform of the smearing function.
We find a non-perturbative expression for $\hat{U}$ in Appendix \ref{appendix:Unitary} that we will use throughout the manuscript. As shown in the appendix, we consider without loss of generality that in the comoving frame $(t,\bm x)$ $t_1\leq t_2\leq \dots \leq t_N$ and 
 obtain: 
\begin{align}
    \label{ProductOfUnis}
    &\hat{U}= \hat{U}_N \hat{U}_{N-1} \dots \hat{U}_1, \\
 \label{ControlUnitary}
     & \hat{U}_\nu =  \sum_{s_\nu}\hat{P}_{s_\nu}\otimes \hat{D}_{s_\nu \beta_\nu} \quad s_\nu = -1,1.   
\end{align}
Here we defined the projector $\hat{P}_{s_\nu}\coloneqq\ket{s_\nu}\!\bra{s_\nu}$, and $\hat{D}_{s_\nu \beta_\nu(\bm{k})}$  are multi-mode displacement operators following the convention in Eq. \eqref{coherent field state}. Note that we can interpret each $\hat{U}_\nu$ as a 'controlled' unitary operator which performs the displacement operator $\hat{D}_{\beta_\nu}$ on the field when the detector $\nu$ is in the state $\ket{1_\nu}$, and its Hermitian conjugate  $\hat{D}_{\beta_\nu}^\dagger=\hat{D}_{-\beta_\nu}$ on the field  when the detector is in the state $\ket{-1_\nu}$.

Applying the unitary $\hat{U}$ from \eqref{ProductOfUnis} on the initial state $\ket{\psi_0}$ we obtain the joint final state, $\ket{\psi}$, for the detector-field system as:
\begin{small}
\begin{align}
\label{psiAllFinal}
    &\ket{\psi} \!=\! \frac{1}{2^{N/2}}\!\sum_{\vec{s}}\exp\!\Big[\ii\sum_{i=0}^N\sum_{j\geq i}^N\! s_js_i \Im(T_{ij})\Big]\!\ket{\vec{s}\,}\! \otimes\! \Big|\sum_{i=0}^N \!s_i\beta_i(\bm{k})\Big\rangle.
\end{align}
\end{small}
Here we have  notated the sum over $\vec{s} \coloneqq s_1,\ldots, s_N$ for a sum over the binary $N$-tuples $(s_1,\dots,s_N)\in \{-1,1\}^N$. The second and third sum are over the indices $i$ which run from $0$ to $N$, and $s_0\coloneqq1$ throughout - to account for the initial state of the field being an arbitrary coherent state of amplitude $\beta_0(\bm{k})$. The final state only depends on pair-wise terms between the detectors and terms coming from the local interaction between the field and each of the detectors. In Eq. \eqref{psiAllFinal} these terms are encoded in the parameters $T_{ij}$, defined as
\begin{align}
\label{Tij2}
  &T_{ij} \coloneqq \frac{\zeta_{ij}}{4}+ \ii\frac{\xi_{ij}}{4} := \int \dd^n\bm{k}\beta_j(\bm{k})\beta^*_i(\bm{k}).
\end{align}
The reader may refer to Appendix \ref{appendix:state} for the detailed derivation.
Tracing over the field in Eq. \eqref{psiAllFinal} we obtain the joint state of the $N$ detectors:

\begin{align}
\label{allDetFinalState}
   \nonumber \hat{\rho}_{\textsc{d}} =& \frac{1}{2^N}\sum_{\vec{s},\vec{s}\,' } \exp[\sum_{i=0}^N T_{ii}(s_is'_i-1)]\\ &\times\exp[\sum_{i,j=0,i>j}^N(T_{ij}s_j-T_{ji}s'_j)(s'_i-s_i)]\big|\vec{s}\,\big\rangle\big\langle\vec{s}\,'\big|.
\end{align}

In the following sections we are going to study correlations between two target detectors. In Appendix \ref{appendix:state} we have derived the reduced density matrix of any pair of detectors. 
For the results of this article, however, we consider the following situation. Alice and Bob control one detector each, labeled A and B. Alice's detector couples to the field before Bob's detector, i.e. $t_\textsc{a} < t_\textsc{b}$. Also, we consider that there are $N$ \textit{interloper} detectors, which couple to the field at times $t_\textsc{a} \leq t_i \leq t_\textsc{b}$ for every $i=1, \ldots, N$.
In that case we obtain a simpler expression particularizing Eq. \eqref{anytwopair}:
\begin{equation}
\label{rho fl}
\begin{split}
    \hat{\rho}_{{\textsc{ab}}}= &\sum_{s_\textsc{a},s_\textsc{b},s'_\textsc{a},s'_\textsc{b}}\Theta(s_\textsc{a},s_\textsc{b},s'_\textsc{a},s'_\textsc{b})  \exp[\ii\,\theta_0(s_\textsc{a},s_\textsc{b},s'_\textsc{a},s'_\textsc{b})] \\
        & \prod_{j\neq \textsc{a},\textsc{b}}\cos{\theta_j(s_\textsc{a},s_\textsc{b},s'_\textsc{a},s'_\textsc{b})} |s_\textsc{a},s_\textsc{b}\rangle\langle s'_\textsc{a},s'_\textsc{b}|.     
\end{split}
\end{equation}
Here
\begin{equation}
\label{rhoflValues1}
    \begin{split}
        \Theta(s_\textsc{a},s_\textsc{b},s'_\textsc{a},s'_\textsc{b}) &= \frac{1}{4}\exp[T_\textsc{bb}(s_\textsc{b}s'_\textsc{b}-1)+T_\textsc{aa}(s_\textsc{a} s'_\textsc{a}-1)] \\
        &\exp[(s_\textsc{b}-s'_\textsc{b})(T_\textsc{ab}s'_\textsc{a}-T_\textsc{ba}s_\textsc{a})] 
    \end{split}
\end{equation}
and 
\begin{align}
\label{thetaVals}
       \theta_0 =\frac{1}{2} (s_\textsc{a}-s'_\textsc{a})&\xi_{0\textsc{a}}+\frac{1}{2}(s_\textsc{b}-s'_\textsc{b})\xi_{0\textsc{b}}, \\
       \theta_j =\frac{1}{2} (s_\textsc{b}-s'_\textsc{b})&\xi_{j\textsc{b}}.\label{Thateq}
\end{align}
In the following section we will show how, even a single \textit{interloper} detector can prevent these two detectors from harvesting correlations.

\section{Cancelling correlations}
\label{result}

As mentioned in the introduction, two initially uncorrelated detectors can become correlated after interacting with a quantum field.  For entanglement in particular, there exists a plethora of results pointing out the different interesting aspects of entanglement harvesting with atomic systems and simplified particle detectors in various scenarios \cite{VALENTINI1991321, Reznik2003-REZEFT,ReznikBenni, steegMeni,brown2013, farming,seismology,genuinetripart, Salton_2015,hydrogenpozas,vacuumEntangleEdu2016,Sitter2017,allison2017,NadineNickLauraAchim}. While amenable  non-perturbatively,
our particular setting (delta-coupled detectors), does not allow for entanglement harvesting due to the no-go theorem in~\cite{nogo}. However,  one can still use delta couplings to non-perturbatively harvest other types of quantum and classical correlations. As we will analyze in Section \ref{sectionTrendsinCorrelation}, with delta couplings it is still possible to harvest correlations from spacelike separated field observables to observables of spacelike separated detectors, harvesting mutual information and quantum discord.

However, in this section we will first present a key result of this paper. We will show how an \textit{interloper} (that is, a adversary that controls a third detector that couples to the field in the temporal past of one of the two detectors harvesting correlations) can completely sabotage the harvesting of any type of correlations between Alice and Bob. 


In this scenario we have three detectors, two target detectors, that is Alice's (A), Bob's (B) (which want to harvest correlations) and the interloper's detector (I), which, in the frame $(t, \bm{x})$, interacts with the field at some time before Bob and after Alice, i.e. $ t_\textsc{a}<t_\textsc{i}<t_\textsc{b}$. To highlight the generality of the result, notice that Alice and Bob can be spacelike separated and hence there may be no frame-independent notion of what interaction happens first.  The density matrix for the target detectors is obtained from~\eqref{allDetFinalState} after tracing out the interloper's detector giving~\eqref{rho fl} that in this particular case is  
\begin{equation}
    \begin{split}
        & \hat{\rho}_{{\textsc{ab}}}=\sum_{s_\textsc{a},s_\textsc{b},s'_\textsc{a},s'_\textsc{b}}|s_\textsc{a},s_\textsc{b}\rangle\langle s'_\textsc{a},s'_\textsc{b}|\Theta(s_\textsc{a},s_\textsc{b},s'_\textsc{a},s'_\textsc{b})\\
        & \exp[\ii\theta_0(s_\textsc{a},s_\textsc{b},s'_\textsc{a},s'_\textsc{b})]\cos{\theta_\textsc{i}(s_\textsc{a},s_\textsc{b},s'_\textsc{a},s'_\textsc{b})},
    \end{split}
\end{equation}
where $\theta_\textsc{i}=\frac{1}{2}(s_\textsc{b}-s'_\textsc{b})\xi_\textsc{ib}$ as per Eq.~\eqref{thetaVals}. $\xi_{\textsc{ib}}$ is a function of the coupling parameters of the interloper and Bob's detectors given by Eq.~\eqref{Tij2}. 

For matrix elements that are non-diagonal for Bob's detector, that is $s_\textsc{b} \neq s'_\textsc{b}$, $\theta_\textsc{i}$ is equal to $\pm \xi_\textsc{ib}$. If we chose $\xi_\textsc{ib}$ to be an odd multiple of $\pi/2$, the corresponding coefficient in $\hat{\rho}_{{\textsc{ab}}}$ vanishes. 
Thus, for this particular choice of $\xi_\textsc{ib}$, the only terms in the density matrix that survive are diagonal in Bob's detector, that is $s_\textsc{b}=s'_\textsc{b}$.
For those elements we have $\Theta = \frac{1}{4}\exp[\frac{\zeta_\textsc{aa}}{4}(s_\textsc{a}s'_\textsc{a}-1)]$, $\theta_\textsc{i}=0$ and $\theta_0=(s_\textsc{a}-s'_\textsc{a})\xi_\textsc{0a}/2$.
This means the matrix elements of $\hat\rho_{\textsc{ab}}$ are independent of the value of $s_\textsc{b}$. Moreover, the density matrix is
\begin{align}
         \hat{\rho}_{\textsc{ab}} =\bigg(\frac{1}{2}\sum_{s_\textsc{a},s'_\textsc{a}}e^{\frac{\zeta_\textsc{aa}}{4}(s_\textsc{a}s'_\textsc{a}-1)+\ii\frac{\xi_\textsc{0a}}{2}(s_\textsc{a}-s'_\textsc{a})}|s_\textsc{a}\rangle\langle s'_\textsc{a}|\bigg)\otimes \frac{1}{2}\openone_\textsc{b}
\end{align}
Thus we see that, when $\xi_\textsc{ib}$ is an odd multiple of $\pi/2$, we have a product state, which in turn implies that there are no correlations whatsoever between Alice and Bob's detectors. In fact, the situation gets even worse for Alice and Bob. By acting on the field in the past lightcone of Bob, the interloper is able to `flood with Bob's detector with entropy; the partial state of Bob's detector is the maximally mixed state. This means that Bob's detector must be maximally entangled with one or more of the other parties involved (because the whole system is in a pure state).

Since the action of tracing out the field following interaction with a simple generated unitary is an entanglement breaking channel, the no-go theorem in~\cite{nogo} tells us that Bob's detector cannot hold any bipartite entanglement with Alice's or the interloper's detector. Bob's detector is however bipartitely entangled with the field, but it is not maximially entangled with it. We know that because there are still correlations between interloper and Bob as we will discuss later in Section~\ref{sectionTheft}. This means---since Bob ends up being maximally mixed---that there has to be genuinely multipartite entanglement between the different parties.

Interestingly, we could understand that the effect of the ``evil'' signal that the interloper sends to Bob modifies the field in such a way that detector B becomes maximally mixed through acquiring bipartite entanglement with the field and sharing multipartite entanglement with the rest of the systems. To illustrate how this is possible, we show a simple example in Appendix~\ref{Appendix:Example}. In sum, we can say that the interaction of the interloper sets the state of the field to saturate Bob's detector with entropy (by creating bipartite and multipartite entanglement that is worthless for the harvesting protocol) when Bob couples.
More importantly, whether Bob's detector ends up in the maximally mixed state only depends on a particular interloper in Bob's chronological past means that adding extra detectors anywhere in spacetime will not prevent this sabotage.
Even if Alice and Bob have an agency of many detectors ready to react to counter the interloper's action through delta-couplings, it would be in vain.
It is conceivable, however, that if the agents are allowed to couple beyond the delta limit that we consider here they may be able to somewhat undo the action of the interloper, but at a complexity cost.
This study is beyond the scope of this paper but should be interesting to explore in the future.

The question of whether a cancellation is possible now boils down to whether there exists an arrangement of detectors for which $\xi_\textsc{ib}=\pi/2$. 
As shown in Appendix \ref{appendix:Tab}, using the Eqs. \eqref{Tab1} and \eqref{Kintegral} we can write,
\begin{align}
\label{xiIB1}
\nonumber
    \xi_\textsc{ib} &= 2\tilde\lambda_\textsc{i}\tilde\lambda_\textsc{b}\int\dd\bm{x'}\dd\bm{z'}F_\textsc{i}(\bm{z'})F_\textsc{b}(\bm{x'}+\bm{z'})\times\\
    &\int_{0}^{\infty}\!\!\dd kJ_{\frac{n}{2}-1}(k\abs{\bm{x'}-\bm{X}})\sin(k T)\bigg(\dfrac{k}{\abs{\bm{x'}-\bm{X}}}\bigg)^{\frac{n}{2}-1}.
\end{align}
Where, $\abs{\bm{X}} = \abs{\bm{x}_\textsc{b} - \bm{x}_\textsc{i}}$ is the spatial separation between the interloper and Bob's detector is  and  $T = t_\textsc{b}-t_\textsc{i}$ is the time delay from the interloper's action to Bob's detector activation.
The smearing function of a localized detector has information about the spatial support of the interaction, and as such, there is a length scale  $\sigma$ providing a scale for the size of the detector. 
A bit more formally, we say that the smearing function is \textit{strongly supported} on a length scale $\sigma$ if $|\bm x|\gg\sigma \Rightarrow F(\bm x)\to 0$. This assumption comes from the locality of the interaction of the detectors and the field: The smearing is zero far away from the spatial support of the detector.

Through a change to dimensionless integration variables in the $k$ integral in Eq.~\eqref{xiIB1} we obtain that in ($n+1$) dimensions
\begin{equation}
    \xi_\textsc{ib} = \frac{\tilde\lambda_\textsc{i}\tilde\lambda_\textsc{b}}{\sigma^{n-1}}\mathcal{I}
\end{equation}
Where, 
\begin{align}
    \nonumber
    \mathcal{I} &= 2\int\dd\bm{x'}\dd\bm{z'}F_\textsc{i}(\bm{z'})F_\textsc{b}(\bm{x'}+\bm{z'})\times\\
    &\int_{0}^{\infty} \dd (k\sigma)J_{\frac{n}{2}-1}(k\abs{\bm{x'}-\bm{X}})\sin(kT)\bigg(\dfrac{k\sigma}{\abs{\bm{x'}-\bm{X}}/\sigma}\bigg)^{\frac{n}{2}-1}.
\end{align}
Importantly, $\mathcal{I}$ is a dimensionless geometric factor that depends on the smearing functions of the detectors and the separations $\abs{\bm{X}}$ and $T$. This factor cannot be very large since the smearing functions are $L^1$ functions normalized to one, so one would expect that this geometric factor is indeed roughly of order 1 for three or less spatial dimensions.

It is important to note that the interloper needs to couple their detector in a region of spacetime that has non-zero overlap with the causal past of Bob in order to sabotage the harvesting protocol and cancel correlations between Alice and Bob. Indeed, the  geometric factor $\mathcal{I}$ is zero unless the interloper's action on the field is (at least partially) in the past lightcone of Bob. We show a proof for the relevant 3+1 dimensional case in Appendix~\ref{appendix:Tab}.  The relative causal separation of the interloper and Alice is irrelevant. In fact if Alice is spacelike separated from Bob which interaction happens first is frame dependent. This tells us that all that the interloper needs to do to sabotage correlations is to couple on the causal past of the detector that they want to flood with entropy and make maximally mixed.
Note that as long as $\mathcal{I}$ is non-zero, we can chose the coupling strengths $\tilde\lambda_\textsc{i}, \tilde\lambda_\textsc{b}$ to be sufficiently large so that $\xi_\textsc{ib}$ can reach the critical value of $\pi/2$. 

In summary, in order for the interloper to completely sabotage the correlations between Alice and Bob, the interloper's coupling strength has to scale as
\begin{equation}
\label{Criteria}
    \tilde\lambda_\textsc{i} \sim \frac{\sigma^{n-1}}{\tilde\lambda_\textsc{b}}.
\end{equation}
Notice that Alice can do very little to protect Bob from this attack. If Bob wants to make it difficult for the interloper to sabotage the protocol and not flood him with entropy, then Bob has to try to couple as nimbly as possible to the field. Of course this also goes in detriment of the amount of correlations that he can achieve with Alice, since that is (at leading order) proportional to $\tilde\lambda_\textsc{a}\tilde\lambda_\textsc{b}$. This means that the safest protocol for correlation extraction is to consider that Alice couples very strongly to the field to allow for the same correlation extraction with a small coupling for Bob, which would make the interloper's life more difficult.

In the above discussion, the number of spacetime dimensions and the choice of smearing functions is arbitrary. Let us study two particular cases to show how easily the interloper can achieve this cancellation in (3+1) dimensions in different scenarios, one with compact support of the smearing functions and one where the smearing functions are taken to be Gaussian.

\subsection{Hard-sphere smearing functions}

First we consider an arbitrary smearing function for Alice's detector and we take Bob's and the interloper's detectors to be localized by an $L^1$ normalised hard sphere in three spatial dimensions:
\begin{equation}\label{hardsphere}
    F_\nu(\bm{x}) =\left\{
        \begin{array}{ll}
            \frac{3}{4\pi\sigma^3} & \quad \abs{\bm{x}} \leq \sigma \\[3mm]
            0 & \quad \abs{\bm{x}} > \sigma
            
        \end{array}
    \right.
\end{equation}
Let us recall that the spatial separation between the interloper and Bob's detector was $\abs{\bm{X}} = \abs{\bm{x}_\textsc{b} - \bm{x}_\textsc{i}}$ and the time delay from the interloper's action to Bob's detector activation $T = t_\textsc{b}-t_\textsc{i}$. In Appendix \ref{appendix:Tab} we show that  
{\small
\begin{equation}
\label{hardsphereq}
\xi_\textsc{ib}\!=\!\begin{cases} 
      0 &\!\!\!\!\frac{|\abs{\bm{X}}- \abs{T}|}{2\sigma}\geq 1 \\[3mm]
     (\frac{1}{5}-\delta_{_-} + \delta_{_-}^{3/2} -\frac{1}{5}\delta_{_-}^{5/2})C &  \frac{\abs{\bm{X}}+\abs{T}}{2\sigma}\geq 1 \\[3mm]
      (\delta_{_+}\!\! - \delta_{_+}^{3/2} +\frac{1}{5}\delta_{_+}^{5/2} -\delta_{_-} \!\!+\! \delta_{_-}^{3/2} -\frac{1}{5}\delta_{_-}^{5/2})C  &  \frac{\abs{\bm{X}}+\abs{T}}{2\sigma}\leq 1
   \end{cases}
\end{equation}
}
with $C \coloneqq \tilde\lambda_\textsc{i}\tilde\lambda_\textsc{b} \frac{3\sqrt{2\pi}}{\sigma\abs{\bm{X}}}$ and $\delta_{_\pm} =\Big(\frac{\abs{\bm{X}}\pm \abs{T} }{2\sigma}\Big)^2$.
To cancel correlations we need to prove that there exist $\abs{\bm{X}} = \abs{\bm{x}_\textsc{b} - \bm{x}_\textsc{i}}$ and $T = t_\textsc{b}-t_\textsc{i}$ such that \mbox{$\xi_\textsc{ib} = (2n+1)\pi/2$} for an integer $n$, given $\sigma$ and $\tilde\lambda_\textsc{b}$.

With simple analysis, it can be shown that the maxima of $\xi_\textsc{ib}$ occurs when $\abs{\bm{X}}=T \approx 0.57673\sigma$. For these values, the condition $\xi_\textsc{ib}\geq\pi/2$ can be written as, 
\begin{equation}
\tilde{\lambda}_\textsc{i} > \frac{\pi\sigma^2}{4\tilde{\lambda}_\textsc{b}}.
\end{equation}
we see that indeed we recover the general scaling law derived from Eq.~\eqref{Criteria}, and that the geometric factor in this case is $\mathcal{I}=2$.

\begin{figure}[t]
\includegraphics[width=0.48\textwidth]{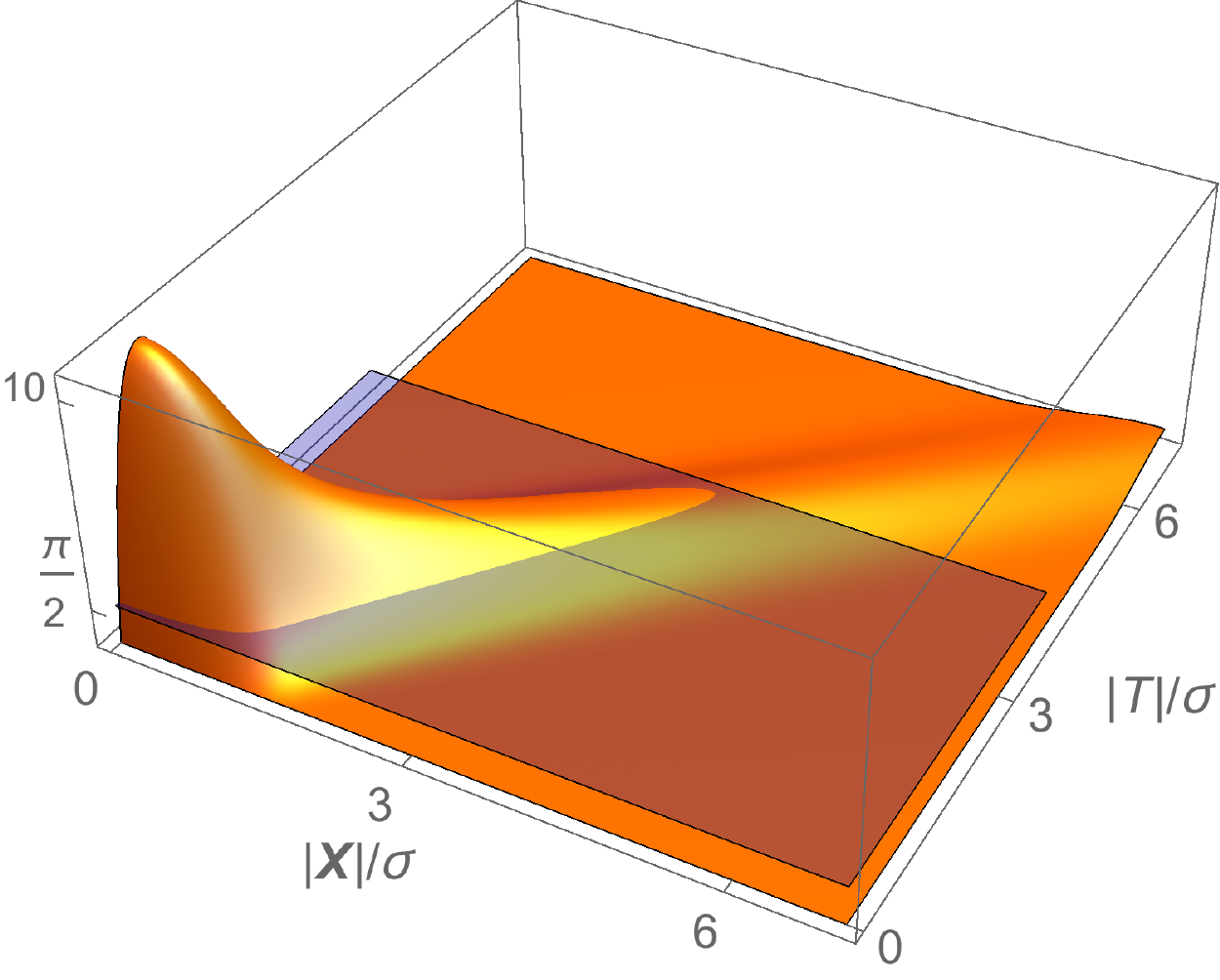}
\caption{Plot showing dependence of $\xi_\textsc{ib}$ on $\abs{\bm{X}} = \abs{\bm{x}_\textsc{b} - \bm{x}_\textsc{i}}$, and  $T = t_\textsc{b}-t_\textsc{i}$. Both interloper and Bob's detectors  have a 3-dimensional hard-sphere smearing function with radius $\sigma$, (Eq. \eqref{hardsphere})  and an effective coupling strength $\tilde\lambda_\textsc{i} = \tilde\lambda_\textsc{b} = 2\sigma$. The intersection of the plane $\xi_\textsc{ib} = \pi/2$  with $\xi_\textsc{ib}(\abs{\bm{X}}, T)$ gives us, for a given location of Bob's detector, possible spacetime positions of the interloper detector for it to sabotage the correlations harvested by Alice and Bob. 
}
\label{HardSpherecan}
\end{figure}

As for where and when the interloper should place the detector, we show in Fig.~\ref{HardSpherecan} a plot with the values of $\xi_{\textsc{ib}}$ as a function of $\abs{\bm{X}}$ and $\abs{T}$ showing also the plane \mbox{$\xi_{\textsc{ib}}=\pi/2$}. The intersection of this plane with $\xi_{\textsc{ib}}(\abs{T},\abs{\bm{X}})$ gives possible spatial and temporal localizations of the interloper detector.

\subsection{Gaussian smearing}
\begin{figure}[t]
\includegraphics[width=0.48\textwidth]{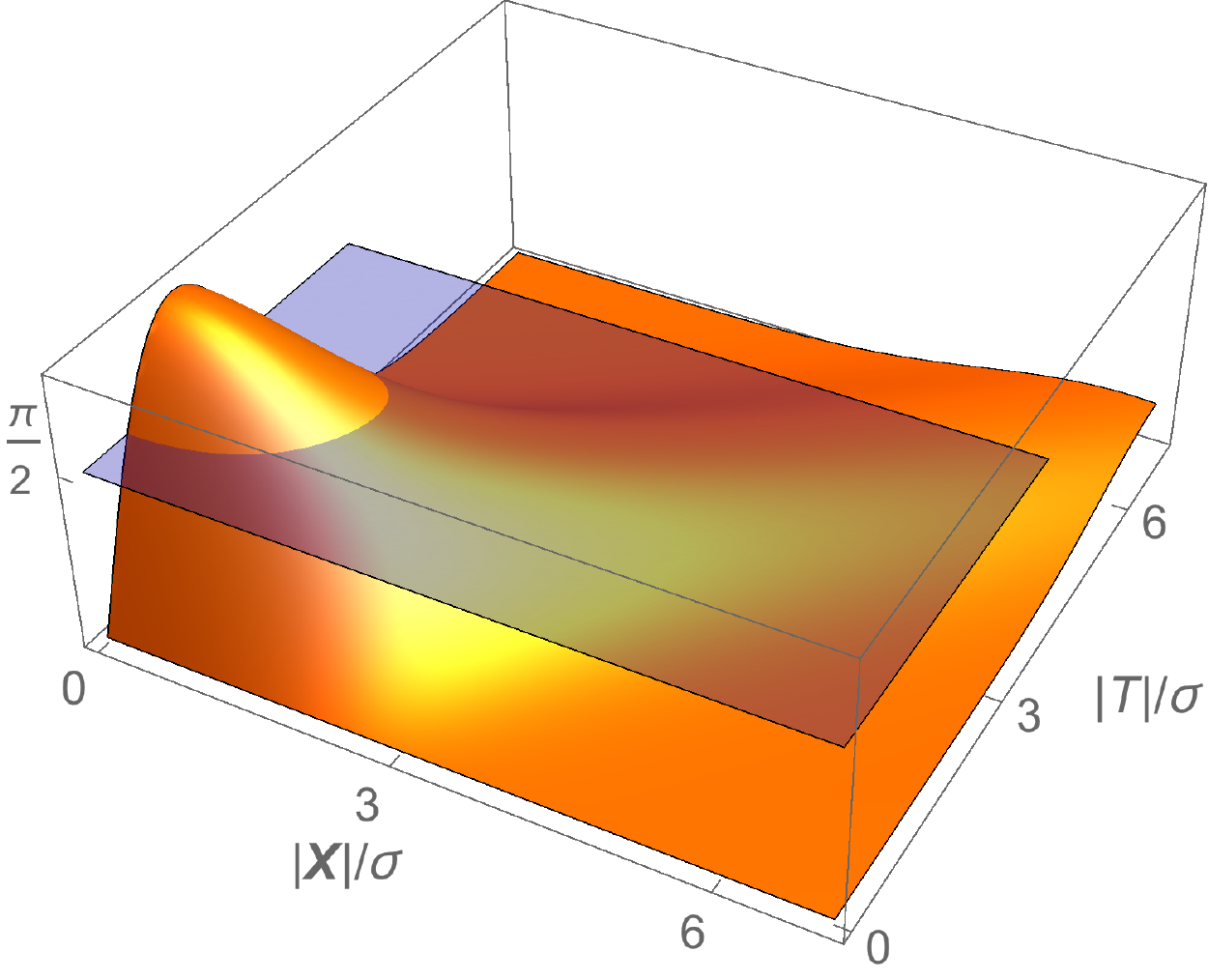}
\caption{Plot showing dependence of $\xi_\textsc{ib}$ on $\abs{\bm{X}} = \abs{\bm{x}_\textsc{b} - \bm{x}_\textsc{i}}$, and  $T = t_\textsc{b}-t_\textsc{i}$. Both interloper and Bob's detectors  have a 3-dimensional Gaussian smearing function with standard deviation $\sigma$, (Eq. \eqref{gaussiansmearing})  and an effective coupling strength $\tilde\lambda_\textsc{i} = \tilde\lambda_\textsc{b} = 2\sigma$. The intersection of the plane $\xi_\textsc{ib} = \pi/2$  with $\xi_\textsc{ib}(\abs{\bm{X}}, T)$ gives us, for a given location of Bob's detector, possible spacetime positions of the interloper detector for it to sabotage the correlations harvested by Alice and Bob. }
\label{QDsthength}
\end{figure}

We now consider the spatial smearing function to be a normalised three-dimensional Gaussian function:
\begin{equation}
\label{gaussiansmearing}
    F_\nu(\bm{x}) = \frac{1}{(2\pi\sigma^2)^{3/2}}\exp[-\frac{\bm{x}^2}{2\sigma^2}].
\end{equation}
In Appendix \ref{appendix:Tab} we obtain the following expression for $\xi_\textsc{ib}$: 
\begin{equation}
    \xi_\textsc{ib}= \frac{\tilde\lambda_\textsc{i}\tilde\lambda_\textsc{b}T}{\sqrt{2}\sigma\abs{\bm{X}}\abs{T}}\Bigg( e^{-\frac{(\abs{T}-\abs{\bm{X}})^2}{4\sigma^2}} - e^{-\frac{(\abs{T}+\abs{\bm{X}})^2}{4\sigma^2}} \Bigg).
\end{equation}
$\xi_\textsc{ib}$ turns out to be a bounded function of $\abs{\bm{X}}$ and $\abs{T}$ which attains a maximum value of $\xi_\textsc{ib}={\tilde\lambda_\textsc{i}\tilde\lambda_\textsc{b}}/(\sigma^2\sqrt{e})$ at $\abs{\bm{X}} = 0$ and $\abs{T}= \sigma\sqrt{2}$.
Again, as the maximum value depends on $\tilde\lambda_\textsc{i}\tilde\lambda_\textsc{b}$, there exists a threshold for the strength of the coupling of the detectors so that $\xi_\textsc{ib}$ exceeds $\pi/2$. The cancellation of correlation becomes possible when
\begin{equation}
    \tilde\lambda_\textsc{i}\geq \frac{\pi\sigma^2\sqrt{e}}{2\tilde\lambda_\textsc{b}}.
\end{equation}
We again see that we recover the general scaling law derived from Eq.~\eqref{Criteria}, and that the geometric factor in this case is $\mathcal{I}=1/\sqrt{e}\approx 0.6$. 
To ascertain possible localizations for interloper's detector we show in Fig.~\ref{QDsthength} a plot with the values of $\xi_{\textsc{ib}}$ as a function of $\abs{\bm{X}}$ and $\abs{T}$ showing also the plane $\xi_{\textsc{ib}}=\pi/2$. The intersection of this plane with $\xi_{\textsc{ib}}(\abs{T},\abs{\bm{X}})$ gives possible spatial and temporal localizations of the interloper detector.

Thus we find that in both the cases, there is a minimum threshold for the coupling strength of the interloper that is controlled by the ratio of the square of the size over the strength of Bob's coupling. We see that the difference in shape only affects the result through geometric factors of order 1. 
We can make the interlopers coupling strength large enough so that the threshold is reached and the cancellation becomes possible.

\section{Correlation measures}
\label{sectionHarvestingCorrelations}

In this section we will study the classical and quantum correlations harvested between two identical target detectors A and B controlled by Alice and Bob respectively, in the presence of other detectors controlled by interlopers, that also couple to the field.
In particular we study three kinds of correlation quantifiers for the final state of the detectors in Eq. \eqref{rho fl}. Namely, we will analyze:
\begin{enumerate}
    \item the correlators of any two arbitrary observables of the detectors.
    \item the mutual information in the state \eqref{rho fl} (that accounts for total correlations acquired by the detectors, both quantum and classical)
    \item the so-called 'classical correlations' (see \cite{henderson2001classical} and discussion in subsection \ref{subsec-Discord}) and the quantum discord in the final two-detector state.
\end{enumerate} 
We shall analyze the dependence of the correlations acquired through the interaction, and harvested from the field on the parameters such as detector coupling strength, relative positioning of detectors and the influence of the presence of additional non-target detectors.

\subsection {Correlators of the detectors Observables}
Let $\hat{\mathcal{O}}_\textsc{a}$ and $\hat{\mathcal{O}}_\textsc{b}$ be any two observables defined on the Hilbert spaces of detectors $A$ and  $B$ respectively.
The correlation function between $\hat{\mathcal{O}}_\textsc{a}$ and $\hat{\mathcal{O}}_\textsc{b}$ in the joint state $\hat{\rho}_{{\textsc{ab}}}$ is defined as
\begin{equation}
\label{corr}
    \Gamma_{\hat \rho_{{\textsc{ab}}}}(\hat{\mathcal{O}}_\textsc{a},\hat{\mathcal{O}}_\textsc{b}) \coloneqq \langle \hat{\mathcal{O}}_\textsc{a}\hat{\mathcal{O}}_\textsc{b}\rangle-\langle\hat{\mathcal{O}}_\textsc{a}\rangle\langle\hat{\mathcal{O}}_\textsc{b}\rangle,
\end{equation}
where $\langle\hat{\mathcal{O}}\rangle\coloneqq \text{Tr}(\hat{\rho}_{{\textsc{ab}}}\hat{\mathcal{O}})$ denotes the expectation value of $\hat{\mathcal{O}}$ on the detectors' state $\hat{\rho}_{{\textsc{ab}}}$ given in Eq. \eqref{rho fl}.

 We define a convenient basis of detector operators for the $\nu-$th detector as 
\begin{equation}
    \begin{split}
        & \hat{S}_0^\nu \coloneqq \ket{-1_\nu}\bra{-1_\nu}+\ket{1_\nu}\bra{1_\nu} = \openone,\\
        & \hat{S}_1^\nu \coloneqq  \ket{-1_\nu}\bra{1_\nu}+\ket{1_\nu}\bra{-1_\nu}, \\
        & \hat{S}_2^\nu \coloneqq  \ii\ket{-1_\nu}\bra{1_\nu}-\ii\ket{1_\nu}\bra{-1_\nu}, \\
        & \hat{S}_3^\nu \coloneqq  -\ket{-1_\nu}\bra{-1_\nu}+\ket{1_\nu}\bra{1_\nu}.
    \end{split}
\end{equation}
where  $\ket{s_\nu}$ are the elements of the basis of each detector's Hilbert space defined in Eq. \eqref{new basis}.
We can then denote any Hermitian operator in the Hilbert spaces of detectors $a$ and $b$ as 
\begin{equation}
    \begin{split}
    & \hat{\mathcal{O}}_\textsc{b} = b_0\hat{S}^\textsc{b}_0 + b_1\hat{S}^\textsc{b}_1 +b_2\hat{S}^\textsc{b}_2 + b_3\hat{S}^\textsc{b}_3, \\
    & \hat{\mathcal{O}}_\textsc{a} = a_0\hat{S}^\textsc{a}_0 + a_1\hat{S}^\textsc{a}_1 +a_2\hat{S}^\textsc{a}_2 + a_3\hat{S}^\textsc{a}_3, 
    \end{split}
\end{equation}
where $a_i, b_i\in \mathbb{R}$.
We recall that the correlation between $\hat{S}_0^\nu=\openone_\nu$ and any operator is zero, thus, we obtain
\begin{equation}
\label{CorrelationOf-Ol}
    \Gamma_{\hat \rho_{{\textsc{ab}}}}(\hat{\mathcal{O}}_\textsc{a},\hat{\mathcal{O}}_\textsc{b}) = \sum_{m,n=1}^3 a_m b_n \Gamma_{\hat \rho_{{\textsc{ab}}}}(\hat{S}_m^\textsc{a},\hat{S}_n^\textsc{b}).
\end{equation}
It is then straightforward to calculate each $\Gamma_{\hat \rho_{\tab}}(\hat{S}_m^\textsc{b},\hat{S}_n^\textsc{a})$ using  \eqref{rho fl}  and \eqref{corr}---which we do explicitly in  Appendix \ref{appendix: correlation}---resulting in
\begin{align}
\label{correlation General}
 \nonumber  
&\Gamma_{\hat \rho_{{\textsc{ab}}}}(\hat{\mathcal{O
   }}_\textsc{a},\hat{\mathcal{O}_\textsc{b}}) =\Big(\prod_{j}\cos{\xi_{j\textsc{b}}}\Big)e^{-\zeta}\\
    \nonumber  
   &\times[\sinh{\zeta_{ab}}(a_1\!\sin{\xi_{0a}}\!-\!a_2\!\cos{\xi_{0a}})(b_1\!\sin{\xi_{0b}}\!-\!b_2\!\cos{\xi_{0b}})\\
    \nonumber  
   &+\!(\cosh{\zeta_{ab}}\!\!-\!\cos{\xi_{ab}}\!)(a_1\!\cos{\xi_{0a}}\!\!+\!a_2\!\sin{\xi_{0a}}\!)b_1\cos{\xi_{0b}}\\
   \nonumber  
   &+\!(\cosh{\zeta_{ab}}\!\!-\!\cos{\xi_{ab}}\!)(a_1\!\cos{\xi_{0a}}\!\!+\!a_2\!\sin{\xi_{0a}}\!)b_2\sin{\xi_{0b}}\\
   &+ e^{\zeta/2}a_3\sin{\xi_{ab}}(b_1\!\sin{\xi_{0b}}\!-\!b_2\!\cos{\xi_{0b}})
   ]
\end{align}
where, since the detectors are considered to be identical and coupling with equal strength, we defined
\begin{equation}
    \zeta\coloneqq \zeta_\textsc{bb}=\zeta_\textsc{aa}=4\int \dd^n \bm k |\beta_\textsc{a}(\bm k)|^2=4\int \dd^n \bm k |\beta_\textsc{b}(\bm k)|^2.
\end{equation}
 As shown in Appendix \ref{appendix:Tab}, $\zeta$ depends on  Alice's and Bob's detectors coupling strengths, but it does not depend on the spacetime position of Alice's, Bob's and the interloper's detectors. 

From Eq. \eqref{correlation General} we can extract how  correlators depend on the parameters of the setup. The first term depends solely on the characteristics of the interloper detectors and their relative position to Bob's detector. The second term, in square brackets, depends on the target detectors and the initial field state.
While it is known that entanglement harvesting from coherent field states is independent from the multi-mode coherence amplitude, $\alpha(\bm{k})$ \cite{nonper}, it is not the case for correlation harvesting. Indeed, the coherent amplitude of the initial field state helps to modulate the correlation harvested, through the parameters $\xi_{0b}$ and $\xi_{0a}$. The role of $\alpha(\bm{k})$ is however restricted only to modulating the correlation between bounds that are decided by terms containing $\zeta_{ab}$, $\xi_{ab}$ and $\zeta$.

\subsection {Mutual information}
Mutual information is a measure of total correlations (classical and quantum) that can be thought of as the amount of information that the two parties in a bipartite system share between each other~\cite{nielsen2000quantum}. It is defined as
\begin{equation}
\label{mutualInfoDef}
    \mathcal{I}(\hat{\rho}_{{\textsc{ab}}})\coloneqq S(\hat{\rho}_\textsc{a})+S(\hat{\rho}_\textsc{b})-S(\hat{\rho}_{{\textsc{ab}}}),
\end{equation}
where $\hat{\rho}_\textsc{a}$ and $\hat{\rho}_\textsc{b}$ are the reduced states for the first and last detector, respectively and $S(\hat{\rho})$ is the von-Neumann entropy of state $\hat{\rho}$.

Although the mutual information can be computed in closed form for any choice of initial coherent state of the field (yielding the result in Appendix \ref{appendix: X-state}), for convenience of the analysis we will focus on the case when the field is initially in the vacuum state. In this scenario the density matrix $\hat{\rho}_{{\textsc{ab}}}$ becomes a $X$-state, as derived in Appendix \ref{appendix: X-state}. 
As shown in the Appendix section~\ref{appendix:mutualInfo}, we obtain 
\begin{align}
\nonumber
    &S(\hat{\rho}_\textsc{a}) = h\Big(e^{-\zeta/2}\cos \xi_\textsc{ab}\prod_{j}\cos{\xi_{j\textsc{b}}}\Big), \\ 
   \nonumber &S(\rho_\textsc{b}) = h(e^{-\zeta/2}), \\
    \nonumber &S(\hat{\rho}_{{\textsc{ab}}}) = -a_1\log_2 a_1 - b_1 \log_2 b_1 + a_1h\Big(\frac{a_2}{a_1}\Big) + b_1 h\Big(\frac{b_2}{b_1}\Big), 
  \end{align}  
 where, 
\begin{align}
	& h(x) = -\frac{1-x}{2}\log_2\Big(\frac{1-x}{2}\Big)-\frac{1+x}{2}\log_2\Big(\frac{1+x}{2}\Big), \nonumber\\
      &a_1 = \frac{1}{2}(1+e^{-\zeta}\cosh \zeta_\textsc{ab}\prod_{j}\cos{\xi_{j\textsc{b}}}), \nonumber\\
    &b_1 = \frac{1}{2}(1-e^{-\zeta}\cosh \zeta_\textsc{ab}\prod_{j}\cos{\xi_{j\textsc{b}}}), \nonumber\\
    &a_2^2 = \frac{1}{4}e^{-\zeta}\bigg[1+2\cos \xi_\textsc{ab}\prod_{j}\cos{\xi_{j\textsc{b}}}  \nonumber\\ 
    &\qquad + \Big(\prod_{j}\cos{\xi_{j\textsc{b}}}\Big)^2\Big(1 +e^{-\zeta}\sinh^2\zeta_\textsc{ab} \Big)\bigg],\nonumber\\
    &b_2^2 = \frac{1}{4}e^{-\zeta}\bigg[1-2\cos \xi_\textsc{ab}\prod_{j}\cos{\xi_{j\textsc{b}}}  \nonumber\\ &\qquad  +\Big(\prod_{j}\cos{\xi_{j\textsc{b}}}\Big)^2\Big(1 +e^{-\zeta}\sinh^2\zeta_\textsc{ab} \Big)\bigg].\label{h(x)}
\end{align}
From the above equations \eqref{h(x)} and the definition \eqref{mutualInfoDef} we can evaluate the mutual information $\mathcal{I}(\hat{\rho}_{{\textsc{ab}}})$.

\subsection {Quantum Discord}
\label{subsec-Discord}
Mutual information does not distinguish classical from quantum correlations.
Even in the absence of entanglement it has been argued that there can be other kind of correlations that may capture some notion of `non-classicality' such as quantum discord~\cite{henderson2001classical,Vedral2003classical,ollivier2001quantum}.
Before defining quantum discord, it is worth introducing  a new  measure of correlations $\mathcal{C}$ (referred to as  \textit{classical correlations} in~\cite{henderson2001classical} but called Henderson-Vedral $\mathcal{C}$ function in this paper for reasons that will become apparent later).
$\mathcal{C}$ is defined as follows:
\begin{equation}
\label{ClassicalCorrDef}
 \mathcal{C}(\hat{\rho}_{{\textsc{ab}}}) = S(\hat{\rho}_\textsc{b}) - \inf_{\{\hat{M}_k\}}S(\hat{\rho}_{{\textsc{ab}}}| \{\hat{M}_k\}),
\end{equation}
where the expression $ S(\hat{\rho}_{{\textsc{ab}}}| \{\hat{M}_k\})$ denotes the average amount of uncertainty that we have about subsystem B after performing local positive-operator valued measures (POVM), $\hat {M}_i\hat {M}_i^\dagger$, on subsystem A:
\begin{align}
\label{EntropyAfterMeasurement}
& S(\hat{\rho}_{{\textsc{ab}}}| \{M_k\}) = \sum_i p_i S(\hat{\rho}_\textsc{b}^i), \\
& p_i = \Tr[(\hat{M}_i\otimes\openone)\hat{\rho}_{{\textsc{ab}}}(\hat{M}_i^\dagger\otimes\openone)], \\
& \hat{\rho}_\textsc{b}^i = \frac{\Tr_\textsc{a}{(\hat{M}_i\otimes\openone)\hat{\rho}_{{\textsc{ab}}}(\hat{M}_i^\dagger\otimes\openone)}}{p_i}.
\end{align}
It can be seen that for a separable state for which the partial state of A is a classical probability distribution over the eigenstates of some observable (i.e., \mbox{$\hat{\rho}_{\textsc{ab}}= \sum_i p_i\ket{i}\!\bra{i} \otimes \hat{\rho}^{i}_\textsc{b}$}), the $\mathcal{C}$ function is equal to the mutual information; $\mathcal{C}(\hat{\rho}_{{\textsc{ab}}})=\mathcal{I}(\hat{\rho}_{{\textsc{ab}}})$~\cite{henderson2001classical}.
Thus, in the case where the system only has classical correlations, $\mathcal{C}(\hat{\rho}_{{\textsc{ab}}})$ is exactly the mutual information. 

However, the presence of entanglement in a pure two-qubit system (which is a genuinely quantum form of correlations) will yield a non-zero $\mathcal{C}$, so a non-vanishing $\mathcal{C}$ does not mean correlations \textit{in absence of entanglement}.
Rather $\mathcal{C}$ accounts for the information that can be learned about B from the application of measurement protocols on A---usually associated with classical (macroscopic) apparatuses acting locally on A.
Along these lines, it can be checked that $\mathcal{C}$ satisfies the following reasonable properties described in \cite{henderson2001classical}:
\begin{itemize}
\item $\mathcal{C} = 0 $ for product states  $\hat{\rho}_{{\textsc{ab}}}=\hat{\rho}_\textsc{a}\otimes\hat{\rho}_\textsc{b}$. 
\item $\mathcal{C}$ is invariant under local unitary transformations.
\item $\mathcal{C}$ is non-increasing under local operations. 
\item $\mathcal{C} = S(\hat{\rho}_\textsc{a})=S(\hat{\rho}_\textsc{b})$ for pure states.
\end{itemize}
The reader can see that the only two differences between the desired properties for $\mathcal{C}$ and an entanglement measure are that a) we allow $\mathcal{C}$ to increase under local operations when there is classical communication and b) we allow $\mathcal{C}$ to be non-zero for non-product separable states.
In this light one may wonder if `classical correlation' might be a misnomer since, for pure bipartite states, $\mathcal{C}$ is exactly the entanglement entropy (which is the paramount measure of entanglement), and entanglement is rarely referred to as a  `classical correlation'.
Instead, the usefulness of $\mathcal{C}$ on its own is that it quantifies the information about B that is revealed when measuring A through POVMs, regardless whether it is a consequence of pre-existing classical correlations or entanglement~\cite{Vedral2003classical}. Note as well that, outside pure states, $\mathcal{C}$ is not symmetric w.r.t subsystems A and B.


The Henderson-Vedral $\mathcal{C}$ function is often difficult to find analytically because of the optimization over the set of possible POVMs on A. 
However, for a pair of qubits  the optimal in~\eqref{ClassicalCorrDef} is always achieved for the smaller set of Projective Value Measurements (PVM) containing two to four elements~\cite{hamierdiscord}. Nevertheless, even in these restricted circumstances, finding the optimal can still be difficult as in general the optimization runs over all possible PVMs (that for a qubit could have up to four elements) and not only von Neumann measurements~\cite{hamierdiscord}. Even though no general analytical expression for $\mathcal{C}$ exists for a pair of qubits in an arbitrary state, some results for certain families of two-qubit states do exist \cite{ali2010quantum,Chen_2011,Galve2011}. For example, in  \cite{Galve2011} it was shown that for rank two density matrices the optimal PVM is indeed an orthogonal measurement and optimizing over von Neumann measurements for density matrices with higher ranks provides a tight bound. 

As for our case, it turns out although the density matrices~\eqref{rho fl}   are not rank two (they have four generally non-zero eigenvalues, indicating it to be Schmidt rank four) so we need to look elsewhere. For our purposes, a particularly relevant family of two-qubit states  are the \textit{X-states} (states for which the density matrix elements in some relevant basis are in the form $\rho_{ij}$, obeying $\rho_{12}=\rho_{13}=\rho_{21}=\rho_{24}=\rho_{31}=\rho_{34}=\rho_{42}=\rho_{43}=0$).
In Appendix \ref{appendix: X-state} we show that when the detectors-field system is initially in its ground state, the joint density matrix $\hat \rho_{\textsc{ab}}$ after the interaction of all the detectors and the field (given by \eqref{rho fl}) is indeed a X-state in the basis $\{\ket{g},e^{\ii t \Omega}\ket{e}\}$. In \cite{ali2010quantum} an algorithm to  analytically evaluate $\mathcal{C}$ for two-qubit \textit{X-states} was proposed, by optimizing over the set of all von Neumann (orthogonal PVMs) measurements. However the algorithm proposed in~\cite{ali2010quantum} does not hold for all two qubit X-states. Indeed a counter example to the algorithm in \cite{ali2010quantum} has been given in \cite{CounterExample}. In more detail, it was shown in \cite{Chen_2011} that, while
the algorithm proposed in~\cite{ali2010quantum} does not hold for all two qubit X-states, it gives the correct results for a certain family of X-states, which satisfy at least one of the following conditions:
\begin{align}
\label{condition1}
    &4(\abs{\rho_{23}}^2)\leq (\rho_{11}-\rho_{22})(\rho_{44}-\rho_{33}),\\
\label{condition2}
    &\abs{\sqrt{\rho_{11}\rho_{44}}-\sqrt{\rho_{22}\rho_{33}}}\leq 2\abs{\rho_{23}}.
\end{align}
In Appendix \ref{appendix: X-state} we show that for $\hat{\rho}_{\textsc{ab}}$ the first inequality (Eq. \eqref{condition1}) is always false but the second condition (Eq. \eqref{condition2}) is always satisfied for the parameters we consider in this manuscript. Hence using the analytical formula provided by Ali et al. in \cite{ali2010quantum} gives the right value for $\mathcal{C}$ justified in our case.
\begin{figure*}
\includegraphics[width=\textwidth]{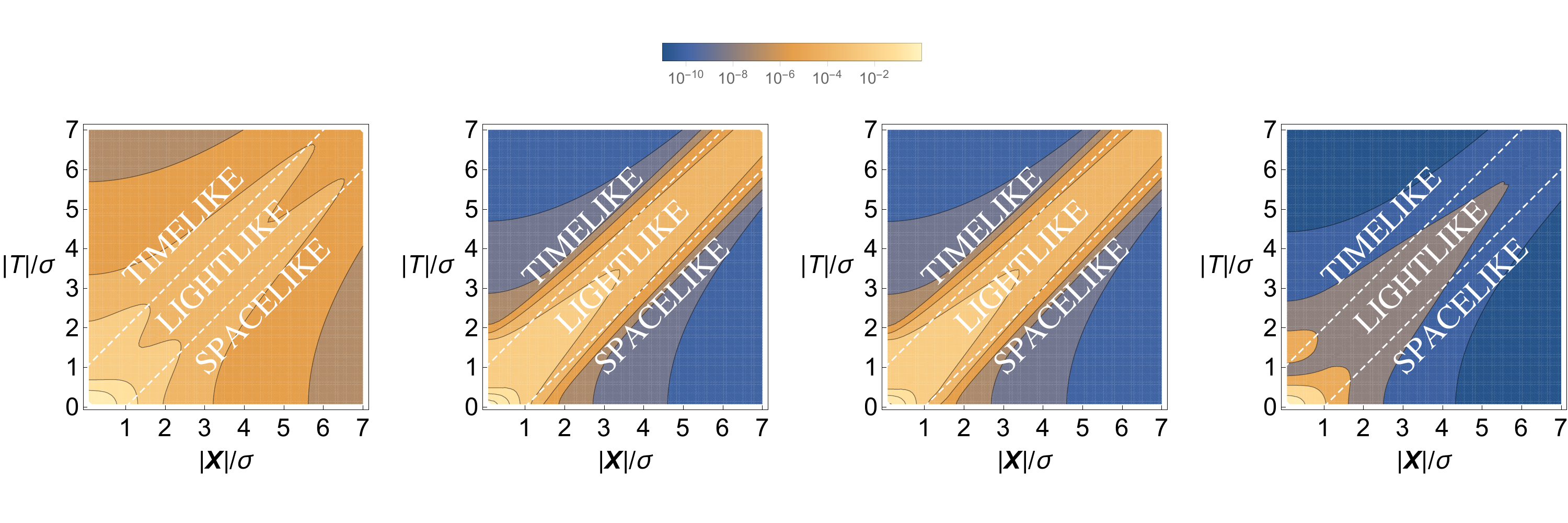}
\begin{tabular}{cccc}
 (a) Energy correlators\:(in units of $\frac{1}{\sigma^2}$)\:\; & (b) Mutual Information \quad  &
\;\; (c) Correlation function $\mathcal{C}$ \quad \quad \quad\:\:& (d) Quantum Discord \quad \quad \\[6pt]
\end{tabular}
\caption{Contour plots showing the dependence of different measures of correlation harvested by Alice and Bob from the field in a vacuum state, on \mbox{$\abs{\bm{X}} = \abs{\bm{x}_\textsc{b} - \bm{x}_\textsc{a}}$, and $T = t_\textsc{b} - t_\textsc{a}$}. Both Alice and Bob's detectors  have a 3-dimensional hard sphere smearing function with radius $\sigma$, (Eq. \eqref{hardsphere}), effective coupling strength $\tilde\lambda_\textsc{i} = \tilde\lambda_\textsc{b} = \sigma$, and energy gaps $\Omega_\ta=\Omega_\tb=1/\sigma$. The axes corresponding to  $\abs{\bm{X}}$ and $T$ are plotted in a linear scale while the different correlation measures are plotted in a logarithmic scale. As the hard sphere detectors are compactly supported, we divide the plots in regions in which Alice and Bob's detectors are completely spacelike separated, lightlike separated and completely timelike separated, revealing that the correlations are strongest when there is some lightlike contact between the detectors.
}
\label{XTdepzoom}
\end{figure*}

\begin{figure*}
\includegraphics[width=\textwidth]{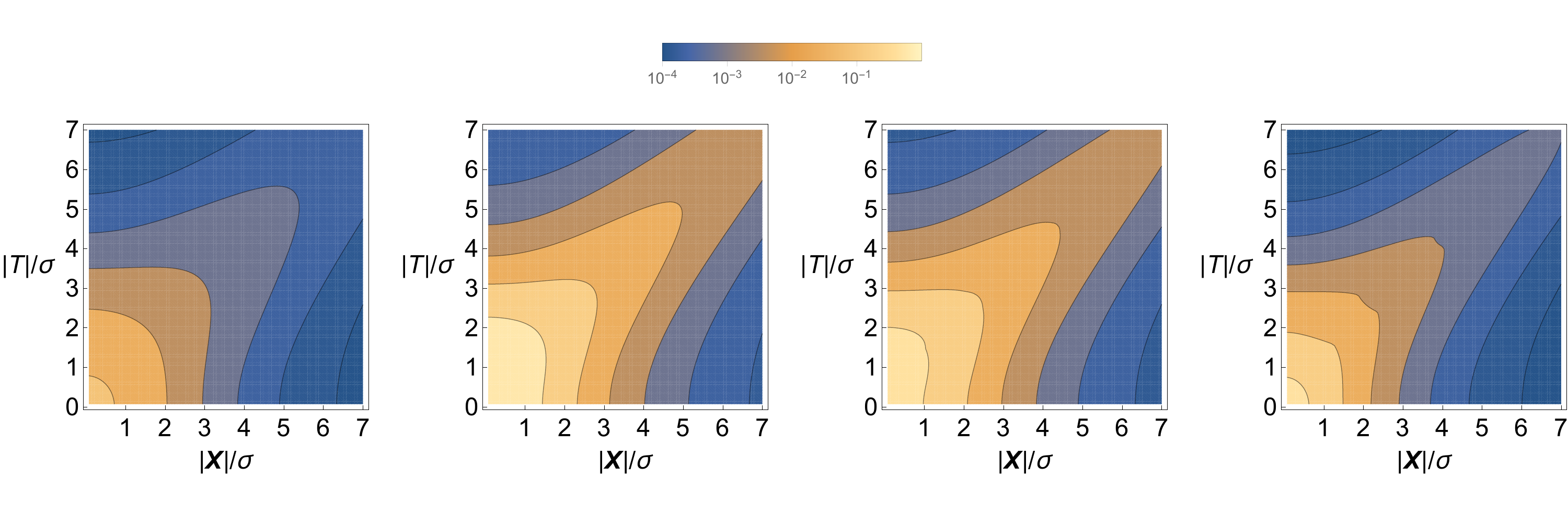}
\begin{tabular}{cccc}
  (a) Energy correlators\:(in units of $\frac{1}{\sigma^2}$)\:\; & (b) Mutual Information \quad  &
\;\; (c) Correlation function $\mathcal{C}$ \quad \quad \quad\:\:& (d) Quantum Discord \quad \quad  \\[6pt]
\end{tabular}
\caption{Contour plots showing the dependence of different measures of correlation harvested by Alice and Bob from the field in a vacuum state, on \mbox{$\abs{\bm{X}} = \abs{\bm{x}_\textsc{b} - \bm{x}_\textsc{a}}$, and $T = t_\textsc{b} - t_\textsc{a}$}. Both interloper and Bob's detectors  have a 3-dimensional hard sphere. smearing function with radius $\sigma$, (Eq. \eqref{gaussiansmearing}), effective coupling strength $\tilde\lambda_\textsc{i} = \tilde\lambda_\textsc{b} = \sigma$, and energy gaps $\Omega_\ta=\Omega_\tb=1/\sigma$. The axes corresponding to  $\abs{\bm{X}}$ and $T$ are plotted in a linear scale while the different correlation measures are plotted in a logarithmic scale.
}
\label{XTdepzoomGauss}
\end{figure*}
To calculate the $\mathcal{C}$-correlation function for a $X$-state as described in \cite{ali2010quantum}, one has to perform an optimization over 8 different possibilities.
In our case, they are reduced to the the following formula as derived in Appendix \ref{appendix: X-state}: 
\begin{align}
\label{CCorrFinal}
    \mathcal{C}(\hat{\rho}_{{\textsc{ab}}}) &= S(\hat{\rho}_\textsc{b}) - \min(p_0 h(\theta_0) + p_1 h(\theta_1), h(\theta')), \\
    p_0 &= \frac{1-e^{-\zeta_\textsc{bb}/2}}{2} \quad \quad p_1 = \frac{1+e^{-\zeta_\textsc{bb}/2}}{2},\\
    \theta_0 &= e^{-\zeta_\textsc{bb}/2} \frac{\cos \xi_\textsc{ba} - e^{-\zeta_\textsc{bb}/2}\cosh \zeta_\textsc{ba} }{2 p_0}\prod_{j}\cos{\xi_{j\textsc{b}}}, \\
    \theta_1 &= e^{-\zeta_\textsc{bb}/2} \frac{\cos \xi_\textsc{ba} + e^{-\zeta_\textsc{bb}/2}\cosh \zeta_\textsc{ba} }{2 p_1}\prod_{j}\cos{\xi_{j\textsc{b}}},  \\
    \theta' &=  e^{-\zeta_\textsc{bb}/2}\sqrt{1+e^{-\zeta_\textsc{bb}}\sinh^2\zeta_\textsc{ba} }\prod_{j}\cos{\xi_{j\textsc{b}}}.
\end{align}

Finally, quantum discord ($\mathcal{Q}$) is a measure of correlations for bipartite systems characterizing quantum correlations that may exist even in the absence of entanglement. 
It is defined as the difference between mutual information, $\mathcal{I(\hat\rho)}$, and the Henderson-Vedral correlation function $\mathcal{C(\hat\rho)}$, 
\begin{equation}
\label{DiscordDef}
    \mathcal{Q}(\hat\rho) = \mathcal{I(\hat\rho)}-\mathcal{C(\hat\rho)}.
\end{equation}
The motivation for defining discord as such comes from the fact that in classical information theory, mutual information between two random variables can be obtained in two equivalent ways: 1) with an expression like \eqref{mutualInfoDef} where the entropies are the respective Shannon entropies, and 2) as an optimization problem over all possible measurements of one of the variables. 
The quantum analogue to the two classically identical expressions can differ for some states. 
Thus discord is defined as the difference of the two~\cite{ollivier2001quantum,henderson2001classical}.

\section{Trends in correlation with and without interlopers }
\label{sectionTrendsinCorrelation}
\begin{figure*}
\includegraphics[width=\textwidth]{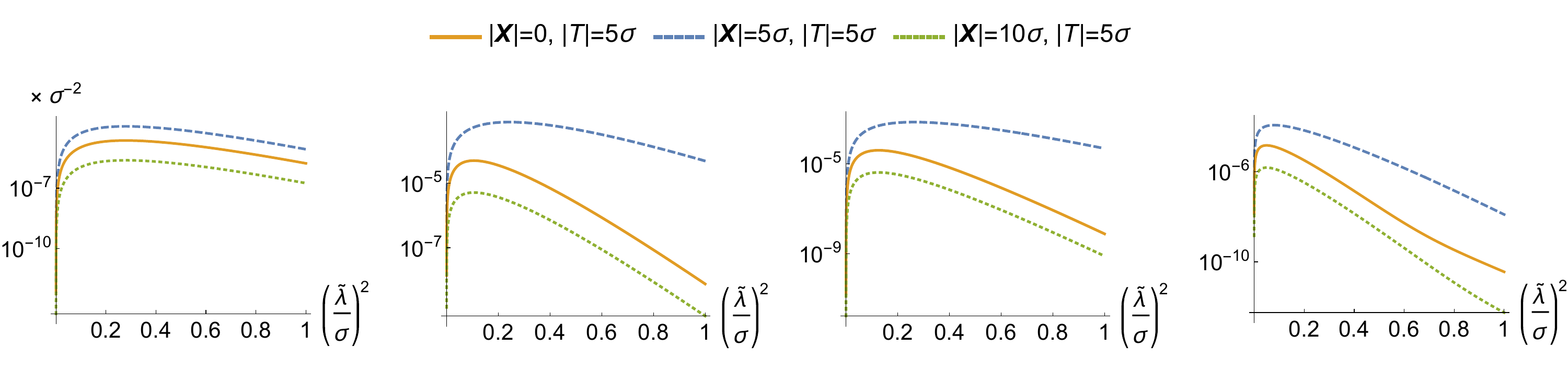}
\begin{tabular}{cccc}
 \quad \quad(a) Energy correlators\:\quad\quad & \quad\quad (b) Mutual Information \quad  & \quad \quad 
\;\; (c) Correlation function $\mathcal{C}$ \quad \quad \quad\quad & (d) Quantum Discord \quad \quad \\[6pt]
\end{tabular}
\caption{Plots showing the dependence of different measures of correlation harvested by Alice and Bob from the field in a vacuum state, in the absence of an interloper, on $\tilde\lambda_\textsc{a}\tilde\lambda_\textsc{b}$, the product of the effective coupling strengths.Both Alice and Bob's detectors have a 3-dimensional hard sphere. smearing function with radius $\sigma$, (Eq. \eqref{hardsphere}), energy gaps $\Omega_\ta=\Omega_\tb=1/\sigma$, and equal effective coupling strength $\tilde\lambda_\textsc{a} = \tilde\lambda_\textsc{b}=\tilde\lambda$. We plot the dependence for three different combinations spacetime separations ($\abs{\bm{X}} = \abs{\bm{x}_\textsc{b} - \bm{x}_\textsc{a}}$, and  $T = t_\textsc{b}-t_\textsc{a}$) between Alice and Bob's interaction with the field: 1) When Alice's and Bob's detectors are completely timelike separated ($\abs{\bm{X}}=0$ and $\abs{T}=5\sigma$ ), 2) When Alice's and Bob's detectors are partly lightlike separated ($\abs{\bm{X}}=5\sigma$ and $\abs{T}=5\sigma$ ) and 3) When Alice's and Bob's detectors are completely spacelike separated ($\abs{\bm{X}}=10$ and $\abs{T}=5$). Note that the axis corresponding to  $\tilde\lambda^2$ is plotted in a linear scale while the different correlation measures are plotted in a logarithmic scale. We note that all correlations vanish when there is no interaction ($\tilde\lambda^2=0$). On increasing the interaction strength the correlations reach an optimal value, before getting attenuated at very high coupling strength.
}
\label{strengthdep}
\end{figure*}
We have seen how the action of the interlopers can always completely cancel the correlations between Alice and Bob. In this section, we will study how the correlations are cancelled as the interloper approaches the optimal sabotage point where all correlations are cancelled. Another interesting question to answer is what happens with the different bipartite correlations in the Alice-Bob-interloper system, and whether the interloper is `stealing' the correlations between Alice and Bob or just merely making them vanish. 

In order to answer these questions, we need to have a baseline with which to compare the results. For this, we study the behaviour of all the different correlation measures for Alice and Bob's detectors in the absence of interlopers, something that, to the authors' knowledge, has not been done in any previous literature.
We will see that quantum and classical correlation harvesting behave similarly in the scenarios considered. In fact, all four correlation measures (Observable correlators, mutual information, the $\mathcal{C}$ correlation function and quantum discord) follow similar trends in their dependence on the parameters of the setup. In this subsection we explore the dependence of the different correlation measures on the coupling strength of the target detectors and their relative positioning.

\subsection{Spacetime dependence of correlations between Alice and Bob}

In the scenario where we have two detectors coupling to the field there are two main ways in which they can get correlated.

First, if the detectors are light-connected, they can talk to each other by the exchange of ``real-quanta''. More precisely,  the first detector creates energy-carrying perturbations that propagate at the speed of light and reach the second detector, correlating the two of them. 

On the other hand, if the detectors are spacelike separated they cannot exchange signals, but they can harvest the correlations that pre-exist in the vacuum state of the field~\cite{VALENTINI1991321, Reznik2003-REZEFT,ReznikBenni, steegMeni,brown2013, seismology,genuinetripart, Salton_2015,hydrogenpozas,vacuumEntangleEdu2016,Sitter2017,allison2017}. This is also the case for pure timelike separation, since in 3+1D Minkowski space, energy carried by a massless field cannot propagate slower than the speed of light due to the strong Huygens principle~\cite{huygensJonsson,mclenaghan1974validity,Czapor}.

In Fig.~\ref{XTdepzoom} we show the dependence of the different measures of correlation as functions of the detectors' relative position for both the hard-sphere and Gaussian smearings. The extracted correlations decay as the space-time separation between the detectors increases.

\subsection{Dependence on the coupling strength}

\begin{figure}[h]
    \includegraphics[scale=0.77]{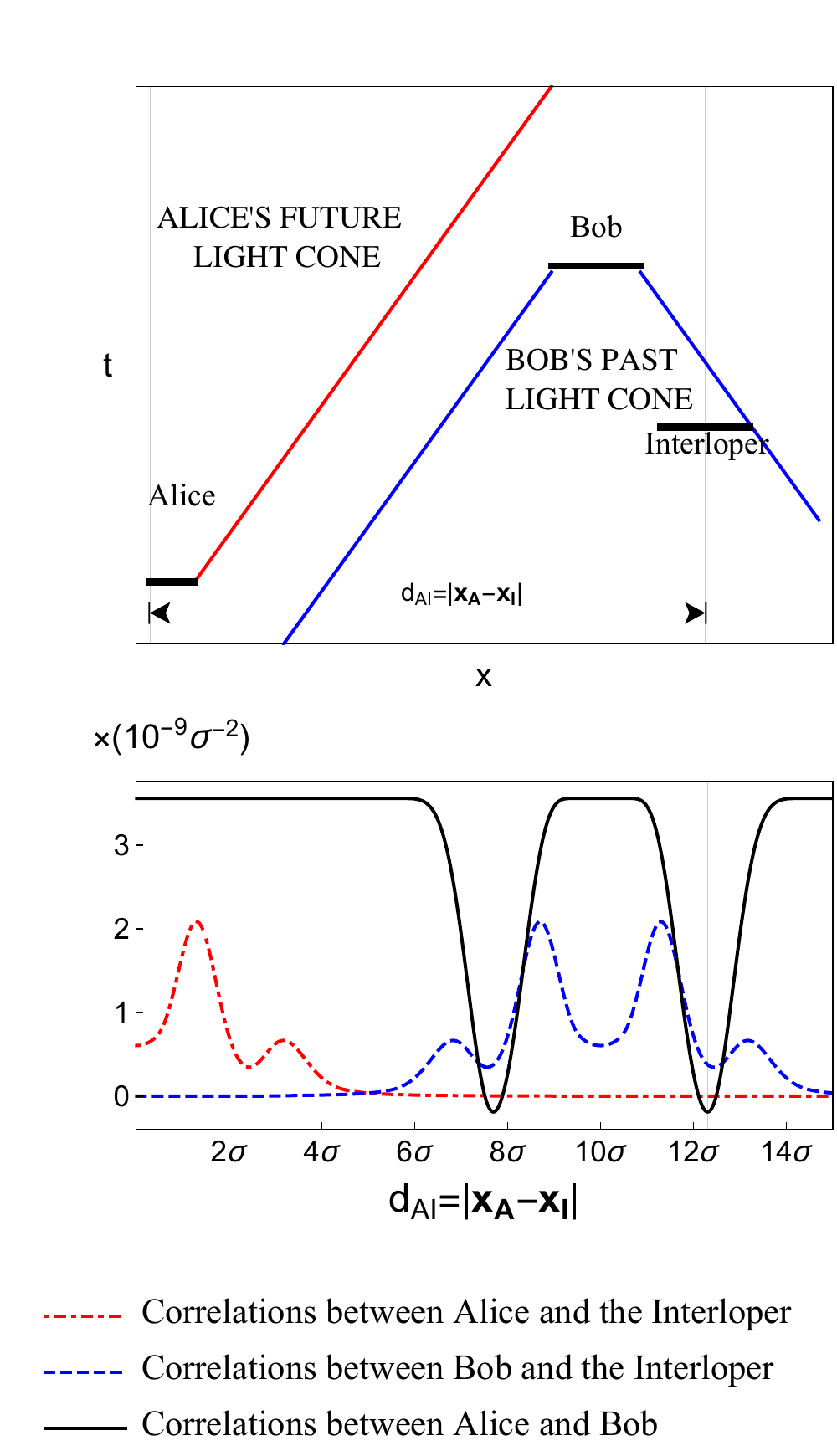}
    \caption{
        We show how the energy correlators between Alice- interloper, Bob-interloper and Alice-Bob in the vacuum state vary when the interloper's center of mass is moved along the line joining Alice and Bob's center of mass, keeping those of Alice and Bob fixed at a spatial separation of $10\sigma$.
        Alice and Bob interact with the field at $t_\ta=0, t_\tb=5\sigma$ respectively, making them completely spacelike separated, and the interloper interacts at the time $t_\ti=2.5\sigma$. 
        We notice that close to the positions where Alice and Bob's harvested correlation is sabotaged, interloper's correlation with Bob also decreases while its correlation with Alice is identically 0, showing that interloper does not steal any correlation from Alice and Bob while sabotaging them.
        The three detectors are characterized by a 3-dimensional hard-sphere smearing function with radius $\sigma$, as given in Eq. \eqref{hardsphere}.
        The energy gap of all detectors is $\Omega_\textsc{a}=\Omega_\textsc{b}=\Omega_\textsc{i}=1/\sigma$. Their coupling strengths are set to $\tilde\lambda_\textsc{a} = \tilde\lambda_\textsc{b}= 1.3\sigma$ and $\tilde\lambda_\textsc{i}= 2\sigma$, respectively.
    }
  \label{CorrelationSchematic}
\end{figure} 
The detectors are most correlated on null-contact (when the detectors can communicate). This happens when\footnote{Not only for the hard-sphere smearing: even if for the Gaussian case there always is some degree of null-contact, the amount of null contact is highly suppressed far from the region $\abs{\abs{\bm{x}_\textsc{b}-\bm{x}_\textsc{a}}-\abs{t_\textsc{b}-t_\textsc{a}}} =\abs{\abs{\bm{X}}-\abs{T}} \leq 2\sigma$.} $\abs{\abs{\bm{x}_\textsc{b}-\bm{x}_\textsc{a}}-\abs{t_\textsc{b}-t_\textsc{a}}} =\abs{\abs{\bm{X}}-\abs{T}} \leq 2\sigma$.
However, even outside of null-contact, we see that detectors can still harvest correlations when they cannot exchange signals, i.e.,  while spacelike or timelike\footnote{Recall that strictly timelike separated detectors cannot signal through a massless field in $3+1$ dimensions Minkowski spacetime due to the strong Huygens principle~\cite{ mclenaghan1974validity,Czapor,huygensJonsson,Mart_n_Mart_nez_2015Caus}} separated. As we discussed before, local field observables in spacelike separation are, in general, also correlated~\cite{Werner1,Werner2,ReznikBenni}, albeit with smaller intensity as the spatial and temporal distance between them increases, and the detectors harvest those correlations.

The strength of the couplings is quantified by the parameter $\tilde\lambda$ (we are assuming the target detectors to be identical; i.e $\tilde\lambda_\textsc{a}=\tilde\lambda_\textsc{b}=\tilde\lambda$ and hence the product of coupling strengths of detectors A and B, is $\tilde\lambda^2$).
In Fig.~\ref{strengthdep} we plot the dependence on $\tilde\lambda^2$ of  (a) the free energy $\hat H_\nu^{\text{free}}$ correlators, (b) mutual information, (c) Henderson-Vedral correlation function and (d) quantum discord.

For small $\tilde\lambda$ the detector correlations increase as the coupling strength increases. The more intensely the detectors couple to the field the more correlations they can acquire. However, there is a critical value of the coupling strength beyond which correlations start to decrease. Indeed, one can prove that in the limit of very strong coupling $\tilde\lambda^2\rightarrow \infty$ all correlations vanish.
To see this we can apply the Cauchy-Schwarz inequality to \eqref{Tij2} obtaining $\abs{\zeta_\textsc{ba}}, \abs{\xi_\textsc{ab}} \leq \zeta$. 
Since by definition, $\zeta\propto\tilde\lambda^2$ and $\zeta\geq 0$, we see that the terms $e^{-\zeta}, e^{-\zeta + \zeta_\textsc{ab}}$ and $e^{-\zeta - \zeta_\textsc{ab}}$ all approach zero as  $\tilde\lambda^2$ tends to infinity.
Thus using Eqs.~\eqref{correlation General},~\eqref{mutualInfoDef},\eqref{h(x)},~\eqref{CCorrFinal}~and~\eqref{DiscordDef} we conclude that all correlation measures must go to zero in the strong coupling limit.

This is expected, and may be explained, as discussed in~\cite{nonper,nogo}, as the fact that the detectors coupling very strongly to the field introduce very strong local noise that overcome the correlations that could be extracted. 
Since the harvesting of correlations is usually a competition between non-local information and local noise~\cite{ReznikBenni}, it is possible that since the strong delta couplings entangle each single detector with the field locally, this local noise drowns out the acquisition of correlations between the two detectors. 

Moreover, as all the correlation measures are continuous at all points, vanish at $\tilde\lambda^2=0$ and asymptote horizontally to the line $y=0$, all correlation measures achieve a maximum for some value of $\tilde\lambda$ (as also shown in Fig.~\ref{strengthdep}). 
Therefore given any arrangement of target detectors, and any correlation measure,  we can always find a coupling strength that maximizes the value of the harvested correlations. 

\subsection{Spacetime positioning of the interlopers and `correlation theft'}
\label{sectionTheft}

In section~\ref{result}, we showed that a single interloper can cancel the harvesting of correlations between Alice and Bob. We proved that, necessarily, the interloper and Bob must be partially light-like connected. When we discussed the interpretation of this result we pictured it as the fact that an ``evil'' placing of a single interloper was modifying the state of the field so that Bob would be ``flooded'' with entropy once it interacts with the field. 

However, one might wonder whether the reason why Bob cannot correlate with Alice is that the interloper is ``stealing'' the correlations that she would have acquired with Bob. In other words, whether by interacting with the field in a particular way at a particular time, the interloper becomes correlated with Alice preventing her from being correlated with Bob or anyone else.

Nevertheless this is not the case at all. In Fig.~\ref{CorrelationSchematic} we vary the position of the interloper moving it towards the point that will maximally sabotage Alice and Bob's attempt to correlate. We see that the correlations between Alice and the interloper decrease monotonically as they get further away at the same time that the correlations between Alice and Bob also decrease. The behaviour of Alice-interloper correlations and Alice-Bob correlations are, perhaps ironically, uncorrelated.

\section{Conclusion}
\label{sectionConclusion}

We have studied non-perturbatively the interactions of an arbitrary number of qubits (modelled as Unruh-DeWitt particle detectors) with an arbitrary coherent state of a scalar field in flat spacetime. The detectors are initialized to their respective ground states and interact with the field via a delta switching function. This commonly employed switching function (see, e.g., \cite{nonper,masahiroEnergyExtraction,Masajo}) allows for non-perturbative studies and models a strong fast interaction of the detectors with the field (faster than any other relevant scale in the problem). The non-perturbative nature of our study allows us to analyze the exact states of the many detectors, a kind of many-body study that is not easily accessible through perturbative tools. 

We focus on the study of the correlations (created through interactions or harvested from the field) acquired between two target detectors operated by Alice and Bob respectively, and how these correlations are affected by presence of other third-party detectors, which we call interlopers. 

First, we find that a single interloper can completely sabotage the acquisition of correlations between the target detectors. This cancellation of correlations can be attained by appropriately placing an interloper in the null past of one of the target qubits. By such a placement of an interloper with a detector configured to sabotage correlations, the joint state of the target detectors becomes a product state $(\rho=\rho_A\otimes\rho_B)$.
Furthermore, the attacked detector (Bob in our case)  becomes maximally mixed, not having any bipartite entanglement with either Alice's or the interloper's detector. We discussed how this cannot be thought of as the interloper `stealing' the correlations between Alice Bob, but rather the interloper setting up the state of the field in a configuration that will `flood' Bob's detector with entropy, preventing it from acquiring correlations with anything else but the field itself. 

It is also possible to place interlopers in positions where their sabotage action is completely independent on any information related to Alice. This would make it impossible for Alice to modify her behaviour to prevent or mitigate the attack. Furthermore, we showed that even if Alice and Bob have help from a coordinated agency of many detectors locally acting on the field, they cannot prevent the sabotage of the acquisition of correlations by the interloper's action.

The question of what should be the exact position of a single interloper to sabotage the correlations, depends on the quantity $\xi_\textsc{ib}$, defined in Eq. \eqref{Tij2},  and whether it can attain the critical value of $\pi/2$.
$\xi_\textsc{ib}$ depends on the relative spatial and temporal distance between the interloper and Bob and the coupling strengths of their respective detectors.
We identify a threshold condition that for a given coupling strength $\tilde{\lambda}_\textsc{b}$ of Bob's detector, the interlopers coupling strength $\tilde{\lambda}_\textsc{i}$ must exceed $\sim \sigma^{n-1}/\tilde{\lambda}_\textsc{b}$, up to an order 1 geometric factor determined by the shape of  of the detectors. Furthermore, we also showed that a swarm of weakly coupled interlopers can achieve the same result.

We notice that to make it difficult for the interloper to sabotage the correlation between Alice and Bob, Bob have to couple with the field very weakly (and thus increasing the threshold $\tilde{\lambda}_\textsc{i}$ inversely). 
However that comes at the cost of correlation values between Alice and Bob being small. 

Although we have studied these scenarios in full generality, we have also considered some illustrative  particular cases. Namely, we have verified the general threshold rule we obtained and solved with full detail the cases where the detectors' smearing functions are compactly supported hard spheres and spherical Gaussians.

To quantify the correlations that the detectors can acquire through interaction or through harvesting from the field we have explicitly evaluated several measures of correlations between Alice and Bob's detector ---The correlators of arbitrary detector observables, mutual information, Henderson-Vedral $\mathcal{C}$ function, and quantum discord. We studied the dependence of these correlation measures on the spacetime location of Alice and Bob, their coupling strengths, and we showed that all them follow similar trends.

Finally, we note that there are experimental setups where the correlations acquired by qubits due to their interaction with a field is detrimental to the purposes of the experiment (see, e.g., ~\cite{Tales}). In those cases, the sabotage of correlations between the two target qubits can be useful to shield the target quantum systems from spurious correlations that would introduce noise in the setup. 

\section*{Acknowledgments}
E.M-M. acknowledges funding  by the NSERC Discovery program as well as his Ontario Early Researcher Award. A.S thanks the USEQIP program at Institute for Quantum Computing, Waterloo that facilitated the beginning of this research. All the authors thank Petar Simidzija for helpful comments on the draft of the paper.

\begin{widetext}

\appendix

\section{Properties of coherent states}
\label{appendix: coherent states}
Recall that a coherent state with coherent amplitude $\alpha(\bm{k})$, $|\alpha(\bm{k})\rangle$, is defined \eqref{coherent field state} as 
\begin{equation}
    |\alpha(\bm{k})\rangle = e^{\big( \int \dd^n \bm{k} \big[ \alpha(\bm{k})\hat{a}_{\bm{k}}^\dagger - \alpha(\bm{k})^*\hat{a}_{\bm{k}}\big ]\big)}\ket{0}.
\end{equation}
The unitary operator that acts on the vacuum state to produce a coherent state is called a displacement operator, 
\begin{equation}
    \hat{D}_{\alpha(\bm{k})} = \exp[\int \dd^n\bm{k} \, (\alpha(\bm{k})\hat{a}_{\bm{k}}^\dagger - \alpha(\bm{k})^*\hat{a}_{\bm{k}})], 
\end{equation}
and it is characterized by a a displacement amplitude distribution $\alpha(\bm{k})$.
The inverse of a displacement operator with amplitude $\alpha(\bm{k})$ is its Hermitian conjugate, which turns out to be another displacement operator with amplitude $-\alpha(\bm{k})$.
Moreover, we know that the creator and annihilator operators satisfy the commutation relation
\begin{equation}
\label{commu1}
   [\hat{a}_{\bm{k}}, \hat{a}^\dagger_{\bm{k'}}] = \delta^{(n)}(\bm{k}-\bm{k'}).
\end{equation}
Using the above commutation relation and Baker–Campbell–Hausdorff identity, we can write the displacement operator as
\begin{equation}
\label{displace}
\begin{split}
      D_{\alpha(\bm{k})}& = \exp[\int \dd^n\bm{k}\alpha(\bm{k})\hat{a}_{\bm{k}}^\dagger]\exp[- \int \dd^n\bm{k} \alpha(\bm{k})^*\hat{a}_{\bm{k}}]\exp[-\frac{1}{2}\int\int \dd^n\bm{k}\dd^n\bm{k'}\delta^{(n)}(\bm{k}-\bm{k'}) \alpha(\bm{k})\alpha^*(\bm{k'})] \\
     & = \exp[\int \dd^n\bm{k}\alpha(\bm{k})\hat{a}_{\bm{k}}^\dagger]\exp[- \int \dd^n\bm{k} \alpha(\bm{k})^*\hat{a}_{\bm{k}}]\exp[-\frac{1}{2}\int \dd^n\bm{k}|\alpha(\bm{k})|^2].
\end{split}
\end{equation}

We want to calculate $\langle0|\alpha_{\bm{k}}\rangle=\langle0|\hat{D}_{\alpha(\bm{k})}\ket{0}$.  In order to do so, we first obtain this intermediate result, using that $\hat{a}_{\bm{k}}$ annihilates the vacuum state:
\begin{equation}
\label{dispac2}
    \exp[- \int \dd^n\bm{k} \alpha(\bm{k})^*\hat{a}_{\bm{k}}]\ket{0} = \Big( \openone - \int \dd^n\bm{k} \alpha(\bm{k})^*\hat{a}_{\bm{k}} + \dots \Big) \ket{0} = \ket{0}.
\end{equation}
Similarly, from the Hermitian conjugate of the above expression, we get
\begin{equation}
\label{dispac3}
    \langle0| = \langle0|\exp[\int \dd^n\bm{k}\alpha(\bm{k})\hat{a}_{\bm{k}}^\dagger].
\end{equation}
From \eqref{displace}, \eqref{dispac2} and \eqref{dispac3} we find the inner product of a coherent state and the vacuum state:
\begin{equation}
\label{ip1}
    \langle0|\alpha_{\bm{k}}\rangle = \langle0|\hat{D}_{\alpha(\bm{k})}\ket{0} = \exp[-\frac{1}{2}\int \dd^n\bm{k}|\alpha(\bm{k})|^2].
\end{equation}
We can also calculate the inner product between two coherent states of coherence amplitudes $\alpha(\bm{k})$ and  $\omega(\bm{k})$, $\langle\alpha(\bm{k})|\omega(\bm{k})\rangle$.Using \eqref{commu1} we can show that 
\begin{equation}
    \label{commu2}
    \Big[\int \dd^n\bm{k} \big[ \alpha(\bm{k})\hat{a}_{\bm{k}}^\dagger - \alpha(\bm{k})^*\hat{a}_{\bm{k}}\big ], \int \dd^n\bm{k} \big[ \omega(\bm{k})\hat{a}_{\bm{k}}^\dagger - \omega(\bm{k})^*\hat{a}_{\bm{k}}\big ]\Big] = \int \dd^n\bm{k} \big(\alpha(\bm{k})\omega^*(\bm{k})- \alpha^*(\bm{k})\omega(\bm{k})).
\end{equation}
Again, using the Baker-Campbell-Hausdorff lemma, we obtain
\begin{equation}
\label{product of displacement ops}
   \hat{D}_{\alpha(\bm{k})}\hat{D}_{\omega(\bm{k})} =  \hat{D}_{\alpha(\bm{k})+\omega(\bm{k})}\exp[\frac{1}{2}\int \dd^n\bm{k} \big(\alpha(\bm{k})\omega^*(\bm{k})- \alpha^*(\bm{k})\omega(\bm{k}))].
\end{equation}
In the above expression, we obtain that a product of two displacement operators is another displacement operator multiplied by a phase factor.
We use \eqref{product of displacement ops} to obtain the inner product of any two arbitrary coherent states, $|\alpha(\bm{k})\rangle$ and $|\omega(\bm{k})\rangle$,
\begin{equation}
    \langle\alpha(\bm{k})|\omega(\bm{k})\rangle = \langle0|\hat{D}_{-\alpha(\bm{k})}\hat{D}_{\omega(\bm{k})}\ket{0}.
\end{equation}
Using \eqref{product of displacement ops} and then \eqref{ip1} we get the inner product of two coherent states.
\begin{equation}
\label{CoherentStateI.P}
    \langle\alpha(\bm{k})|\omega(\bm{k})\rangle =
    \exp[\frac{-1}{2}\int \dd^n\bm{k}\big( |\alpha(\bm{k})|^2 + |\omega(\bm{k})|^2 - 2\alpha^*(\bm{k})\omega(\bm{k}))].
\end{equation}

\section{Calculating non-perturbative time evolution}
\label{appendix:Unitary}

Starting from equation \eqref{UnitaryDef} in the main text  and using the time switching functions $\chi_\nu(t)=\eta_\nu\delta(t-t_\nu)$ we have

\begin{align}
\nonumber
   \hat{U} &= \mathcal{T}\exp[-\ii\int_{-\infty}^\infty \!\!\!\dd t \hat{H}(t)] = \mathcal{T} \exp[-\ii\sum_{\nu=1}^N\hat{H}_\nu] \\
   \nonumber
    &=\sum_{n=0}^\infty \frac{1}{n!} \sum_{m_i, \sum m_i = n} \frac{n!}{m_1! m_2!\dots m_N!}(-\ii\hat{H}_N)^{m_N} (-\ii\hat{H}_{N-1})^{m_{N-1}}\dots(-\ii\hat{H}_1)^{m_1} \\
    \nonumber
    &= \left( \sum_{m_N=0}^\infty\frac{1}{m_N!} (-\ii\hat{H}_N(t_N))^{m_N}) \right) \left( \sum_{m_{N-1}=0}^\infty\frac{1}{m_{N-1}!} (-\ii\hat{H}_{N-1}(t_{N-1}))^{m_{N-1}}) \right)\dots \left( \sum_{m_1=0}^\infty \frac{1}{m_1!}(-\ii\hat{H}_1(t_1))^{m_1}) \right)\\
    &= \hat{U}_N \hat{U}_{N-1} \dots \hat{U}_1, 
\end{align}
where  $\hat{H}_\nu$ are defined in Eq. \eqref{Hnu} as $\hat{H}_\nu = \lambda_\nu\hat{\mu}_\nu(t_\nu)\otimes\int \dd^n\bm{x} F_\nu(\bm{x}-\bm{x}_\nu)\hat{\phi}(t_\nu,\bm{x})$. Also we have defined $\hat{U}_\nu = \exp[-\ii\hat{H}_\nu]$. 
Note that we have assumed that $t_1\leq t_2 \leq \dots \leq t_N$ to perform the time ordering. Plugging in the expression of the field from Eq. \eqref{field} we get

\begin{equation}
\begin{split}
    \hat{H}_\nu = \lambda_\nu\hat{\mu}_\nu(t_\nu)\otimes\int \dd^n\bm{x} F_\nu(\bm{x}-\bm{x}_\nu)\int \dd^n\bm{k} \frac{1}{\sqrt{2(2\pi)^n|\bm{k}|}}\Big[ \hat{a}_{\bm{k}}^\dagger e^{\ii(|\bm{k}|t_\nu - \bm{k}\cdot\bm{x})}+ \hat{a}_{\bm{k}} e^{-\ii(|\bm{k}|t_\nu - \bm{k}\cdot\bm{x})} \Big].
\end{split}
\end{equation}
Using the Fourier transformation of $F_\nu(\bm{x}-\bm{x}_\nu)$ from Eq. \eqref{fourierTransform}, we can perform the integration over $\bm{x}$ in the above expression to obtain

\begin{equation}
    \hat{H}_\nu = \lambda_\nu\hat{\mu}_\nu(t_\nu)\otimes\int \dd^n\bm{k} \frac{1}{\sqrt{2|\bm{k}|}}\Big[\Tilde{F_\nu}(-\bm{k}) \hat{a}_{\bm{k}}^\dagger e^{\ii(|\bm{k}|t_\nu - \bm{k}\cdot\bm{x}_\nu)}+ \Tilde{F_\nu}(\bm{k}) \hat{a}_{\bm{k}} e^{-\ii(|\bm{k}|t_\nu - \bm{k}\cdot\bm{x}_\nu)} \Big].
\end{equation}
Using $\beta_\nu(\bm{k})$ from equation \eqref{betaNuDef}, we may now rewrite $\hat{U}_\nu$ as
\begin{equation}
    \hat{U}_\nu = \exp[\hat{\mu}_\nu\otimes  \int \dd^n \bm{k} \big[ \beta_\nu(\bm{k})\hat{a}_{\bm{k}}^\dagger - \beta_\nu(\bm{k})^*\hat{a}_{\bm{k}}\big ]].
\end{equation}
Observe that $\hat{\mu}_\nu^2=\openone$. Keeping this in mind, we may expand the exponential in $\hat{U}_\nu$ to obtain
\begin{equation}
    \hat{U}_\nu= \frac{1+\hat{\mu}_\nu}{2}\otimes \hat{D}_{\beta_\nu(\bm{k})}+\frac{1-\hat{\mu}_\nu}{2}\otimes \hat{D}_{-\beta_\nu(\bm{k})}.
\end{equation}
In the $\{\ket{s_\nu}\}$ basis, as given in \eqref{new basis}, it is easy to see that
\begin{equation}
  \begin{split}
      &\hat{m}_\nu(t_\nu) = |1_\nu\rangle\langle1_\nu| - |-1_\nu\rangle\langle-1_\nu| = \hat{S}_3^\nu, \\
     & \hat{\mu}_\nu(t_\nu) = \openone_1\otimes\dots\otimes\openone_{\nu-1}\otimes\hat{S}_3^\nu\otimes\dots\openone_N.\\
  \end{split}   
\end{equation}
Using this in the expression for $\hat{U}_\nu$ we obtain equation \eqref{ControlUnitary},
\begin{equation}
  \hat{U}_\nu =   \sum_{s_\nu}\hat{P}_{s_\nu}\otimes\hat{D}_{s_\nu\beta_\nu(\bm{k})}, \quad s_\nu = -1,1  .
\end{equation}
Here, $\hat{P}_{s_\nu} = \openone_1\otimes\dots\otimes\openone_{\nu-1}\otimes\ket{s_\nu}\bra{s_\nu}\otimes\dots\openone_N$ is the projector on to the $\ket{s_\nu}$ subspace. 
The unitary operator for the entire interaction $\hat{U}_\nu$ is thus a product of the individual unitaries $\hat{U}_\nu$ as obtained in equation \eqref{ProductOfUnis}.
It can be further written as
\begin{equation}
    \hat{U}= \hat{U}_N\hat{U}_{N-1}\dots\hat{U}_1 =\sum_{\{s_1,s_2,\dots s_N\}}\hat{P}_{s_N}\dots\hat{P}_{s_1} \otimes \hat{D}_{s_N\beta_N(\bm{k})}\dots \hat{D}_{s_1\beta_1(\bm{k})}.
\end{equation}
The summation is over all the possible $N$-tuples $\{s_1,s_2,\dots,s_N\}$, where each $s_\nu$ takes the value 1 or -1.
In equation \eqref{product of displacement ops}, we obtain an expression for the product of two displacement operators. 
Using it repeatedly, we can obtain the product of \textit{N} displacement operators,
\begin{equation}
\label{product of exp}
    \hat{D}_{s_N\beta_N(\bm{k})}\dots \hat{D}_{s_1\beta_1(\bm{k})} = \hat{D}_{s_N\beta_N(\bm{k})+\dots+s_1\beta_1(\bm{k})} \exp[\ii\Im\sum_{j\geq i}s_js_iT_{ij}],
\end{equation}
where
\begin{equation}
    \label{Tij}
    T_{ij}= \frac{\zeta_{ij}}{4}+ \ii\frac{\xi_{ij}}{4} := \int \dd^n\bm{k}\beta_j(\bm{k})\beta^*_i(\bm{k}).
\end{equation}
The evaluation of $T_{ij}$ depends on the choice of smearing function. We provide an expression for $\xi_{ij}$ and $\zeta_{ij}$ for a hard-sphere smearing function and a Gaussian smearing function in Appendix \ref{appendix:Tab}.

\section{Calculation of the final state of the detectors}
\label{appendix:state}
In this section we calculate the joint state of any two detectors (labelled as $j_\ta, j_\tb \in \{1,N\}$) out of $N$ detectors after the interaction of the $N$ detectors with the field. 
Initially all the detectors are in their respective ground state, $\ket{g_\nu}$, and the field in a coherent state, $\ket{\beta_0(\bm{k})}$, defined in Eq. \eqref{coherent field state}.
We can write the initial state of all the detectors and the field in the $\ket{s_\nu}$ basis, as given in \eqref{new basis}, as
\begin{equation}
    \ket{\psi_0}= \frac{1}{\sqrt{2}}\big(\ket{1_1}+\ket{-1_1}\big)\otimes\dots\otimes\frac{1}{\sqrt{2}}\big(\ket{1_N}+\ket{-1_N}\big)\otimes\ket{\beta_0(\bm{k})} = \frac{1}{\sqrt{2^N}}\sum_{\vec{s'}}\bigket{\vec{s'}\,} \otimes \hat{D}_{\beta_0(\bm{k})}\bigket{0}.
\end{equation}
Here we have  notated the sum over $\vec{s} \coloneqq s_1,\ldots, s_N$ for a sum over the binary $N$-tuples $(s_1,\dots,s_N)\in \{-1,1\}^N$.
The final state of the detector and the fields thus is,
\begin{equation}
\begin{split}
   \ket{\psi} = \hat{U}\ket{\psi_0} &=
    \frac{1}{\sqrt{2^N}}\sum_{\vec{s}}\hat{P}_{s_N}\dots\hat{P}_{s_1} \otimes \hat{D}_{s_N\beta_N(\bm{k})}\dots \hat{D}_{s_1\beta_1(\bm{k})} \sum_{\vec{s}\,'} \bigket{\vec{s}\,} \otimes \hat{D}_{\beta_0(\bm{k})} \bigket{0} \\
    & = \frac{1}{\sqrt{2^N}}\sum_{\vec{s}\,'} \bigket{\vec{s}\,} \otimes \hat{D}_{s_N\beta_N(\bm{k})}\dots \hat{D}_{s_1\beta_1(\bm{k})}\hat{D}_{\beta_0(\bm{k})}\bigket{0}.
\end{split}
\end{equation}
Using equation \eqref{product of exp} to evaluate the product of the displacement operators, we retrieve Eq. \eqref{psiAllFinal},

\begin{equation}
\ket{\psi}=  \frac{1}{\sqrt{2^N}}\sum_{\vec{s}}\exp[\ii\sum_{i=0}^N\sum_{j\geq i}^N s_js_i \Im(T_{ij})] \bigket{\vec{s}\,} \otimes \Bigket{\sum_{i=0}s_i\beta_i(\bm{k})}.
\end{equation}
As explained in the text, the first sum is over the possible sets $\{s_1,\dots,s_N\}$, the second and third sum are over the indices \textit{i} which run from 0 to N, with the understanding that $s_0=1$.
The ket $\ket{\sum_{i=0}s_i\beta_i(\bm{k})}$ denotes a coherent state with coherent amplitude $\sum_{i=0}s_i\beta_i(\bm{k})$ and $\ket{\vec{s}}$ denotes the joint state of N detectors $\ket{s_1,s_2,\dots,s_N}$ as defined above.
Now that we have found the final joint detector and field state, we can trace out the field degree of freedom to obtain the state of the detectors,
\begin{equation}
\label{detector state intmdt}
     \hat{\rho}= \Tr_{\phih}[\ket{\psi}\bra{\psi}] = \frac{1}{2^N}\sum_{\vec{s},\vec{s}'}\bigket{\vec{s}\,}\bigbra{\vec{s}\,'} \exp[\ii\sum_{i=0}^N\sum_{j\geq i}^N s_js_i \Im(T_{ij})]\exp[-\ii\sum_{i=0}^N\sum_{j\geq i}^N s'_js'_i \Im(T_{ij})] 
     \Bigbra{\sum_{i=0}s'_i\beta_i(\bm{k})}\sum_{i=0}s_i\beta_i(\bm{k})\Big\rangle.
\end{equation}
We use Eq. \eqref{CoherentStateI.P} to calculate the inner product of the coherent states in Eq. \eqref{detector state intmdt}. 
It turns out to be
\begin{equation}
\label{inner product}
    \Big\langle\sum_{i=0}s'_i\beta_i(\bm{k})\Big|\sum_{i=0}s_i\beta_i(\bm{k})\Big\rangle  = \exp[ -\frac{1}{2}\sum_{i=0,j=0}^NT_{ij}(s_is_j + s'_is'_j-2s'_is_j)].
\end{equation}
Using the above expression in Eq. \eqref{detector state intmdt}, and after performing some algebraic simplifications, we obtain the state of all the detectors in Eq. \eqref{allDetFinalState} as
\begin{equation}
    \hat{\rho}_\textsc{d} = \frac{1}{2^N}\sum_{\vec{s},\vec{s}'}\bigket{\vec{s}\,}\bigbra{\vec{s}\,'} \exp[\sum_{i=0}^N T_{ii}(s_is'_i-1) +\sum_{i=0}^N\sum_{j<i}^N(T_{ij}s_j-T_{ji}s'_j)(s'_i-s_i)].
\end{equation}
The joint state of any two detectors ($j_\ta$ and $j_\tb$)is readily obtained from \eqref{allDetFinalState} by tracing over all other detectors \textit{i}, such that $i \in \{1,2,\dots,N\}$ and $i\neq j_\textsc{a}, i\neq j_\textsc{b}$.
Without loss of generality, let's assume that $j_\ta<j_\tb$. Performing the partial trace, we obtain 
\begin{equation}
\label{anytwopair}
    \hat{\rho}^{\textsc{ab}}=\sum_{s_{j_\ta},s_{j_\tb},s'_{j_\ta},s'_{j_\tb}}|s_{j_\ta},s_{j_\tb}\rangle\langle s'_{j_\ta},s'_{j_\tb}|\Theta(s_{j_\ta},s_{j_\tb},s'_{j_\ta},s'_{j_\tb})\exp[\ii\theta_0(s_{j_\ta},s_{j_\tb},s'_{j_\ta},s'_{j_\tb})]\prod^N_{j\neq j_\ta,j_\tb,j=1}\cos{\theta_j(s_{j_\ta},s_{j_\tb},s'_{j_\ta},s'_{j_\tb})},
\end{equation}
where,

\begin{align}
\label{anyTwoPairVals}
\begin{split}
        &\Theta(s_{j_\ta},s_{j_\tb},s'_{j_\ta},s'_{j_\tb})= \frac{1}{4}\exp[T_{j_\ta j_\ta}(s_{j_\ta}s'_{j_\ta}-1)+T_{j_\tb j_\tb}(s_{j_\tb}s'_{j_\tb}-1)+(s_{j_\tb}-s'_{j_\tb})(T_{j_\ta j_\tb}s'_{j_\ta}-T_{j_\tb j_\ta}s_{j_\ta})], \\
        &\theta_j(s_{j_\ta},s_{j_\tb},s'_{j_\ta},s'_{j_\tb})
            = \left \{
            \begin{array}{lll}
              2(s_{j_\tb}-s'_{j_\tb})\Im[T_{j{j_\tb}}]+2(s_{j_\ta}-s'_{j_\ta})\Im[T_{j j_\ta}] & \quad \forall \; 0\leq j<{j_\ta}   \\
                2(s_{j_\tb}-s'_{j_\tb})\Im[T_{j{j_\tb}}] & \quad \forall \; {j_\ta}<j<{j_\tb} \\
                0 & \quad \forall \; j>{j_\tb}\\
           \end{array}
           \right \}.
\end{split}
\end{align}
In this paper we only consider the first and the last detector as our targets for correlation harvesting.
Thus we consider a particular case of the above, where Alice's is the first detector($j_\ta=1$) to interact with the field  and Bob's is the last detector ($j_\ta=N+2$) to interact with the field, with \textit{N} detectors ($j \in \{2,\dots, N+1\}$)interacting in between.
The state of the Alice and Bob's detectors given in \eqref{rho fl} is then readily obtained from Eqs. \eqref{anytwopair} and \eqref{anyTwoPairVals}.
For simplicity now we relabel the interloper detectors from $1$ to $N$ and label Alice and Bob's detector simply as A and B respectively.
The density matrix of the target detectors \eqref{rho fl} written in the basis $\{\ket{-1_\textsc{a}}\ket{-1_\textsc{b}},\ket{-1_\textsc{a}}\ket{1_\textsc{b}},\ket{1_\textsc{a}}\ket{-1_\textsc{b}},\ket{1_\textsc{a}}\ket{1_\textsc{b}}\}$ is given as follows:

\begin{equation}
\label{rhoMatrixFormSbasis}
\rho_{\textsc{ab}}= \frac{1}{4}
\left(\begin{array}{cccc}
 1 & \rho_{12} & \rho_{13} &\rho_{14} \\
 \rho_{12}^* & 1 & \rho_{23} &\rho_{24}\\
 \rho_{13}^* & \rho_{23}^* & 1 & \rho_{34}
   \\
 \rho_{14}^* & \rho_{24}^* & \rho_{34}^* &
   1 \\
\end{array}
\right),
\end{equation}
where we consider $\zeta \coloneqq \zeta_\textsc{bb}= \zeta_\textsc{aa} $ and have the following matrix entries: 

\begin{equation}
\label{rho matrix values S basis}
    \begin{split}
        & \rho_{12}=  \exp[-\frac{\zeta}{2}-\ii \xi_{0\textsc{b}}+\ii\xi_{\textsc{ab}}]\prod_{j=1}^N\cos (\xi_{j\textsc{b}}), \quad \quad \rho_{13} =   \exp[-\frac{\zeta}{2}-\ii \xi_{0\textsc{a}}]  , \quad \quad \rho_{14}=  \exp[- \zeta - \zeta_{\textsc{ab}}-\ii \xi_{0\textsc{a}}-\ii \xi_{0\textsc{b}}] \prod_{j=1}^N\cos (\xi_{j\textsc{b}}),\\
        &\rho_{23}= \exp[ - \zeta+\zeta_{\textsc{ab}}+\ii \xi_{0\textsc{b}}-\ii \xi_{0\textsc{a}}] \prod_{j=1}^N\cos (\xi_{j\textsc{b}}), \quad \quad \rho_{24} = \exp[-\frac{\zeta}{2}-\ii \xi_{0\textsc{a}}],  \quad \quad \rho_{34}=\exp[-\frac{\zeta}{2} - \ii\xi_{0\textsc{b}}-\ii\xi_{\textsc{ab}}]\prod_{j=1}^N\cos (\xi_{j\textsc{b}})
    \end{split}
\end{equation}

\section{Calculating observable correlations}
\label{appendix: correlation}

Here we calculate the correlations $\Gamma(\hat{S}_m^\textsc{a} , \hat{S}_n^\textsc{b})$ for $m,n \in \{1,2,3\}$. We can write the operators $\hat{S}^\nu_i$ in the $\{\ket{s_\nu}\}$ basis.
For example,\begin{align}
&  \hat{S}_1^\textsc{a} = (\ket{-1_\textsc{a}}\bra{1_\textsc{a}}+\ket{1_\textsc{a}}\bra{-1_\textsc{a}}) \otimes   \openone_\textsc{b}  =
\left(    \begin{array}{cccc}
         0 & 0 & 1 & 0  \\
         0 & 0 & 0 & 1 \\
         1 & 0 & 0 & 0 \\
         0 & 1 & 0 & 0 \\
    \end{array}
\right), \\
&   \hat{S}_1^\textsc{b} =  \openone_\textsc{a}  \otimes (\ket{-1_\textsc{b}}\bra{1_\textsc{b}}+\ket{1_\textsc{b}}\bra{-1_\textsc{b}}) =
\left(    \begin{array}{cccc}
         0 & 1 & 0 & 0  \\
         1 & 0 & 0 & 0 \\
         0 & 0 & 0 & 1 \\
         0 & 0 & 1 & 0 \\
    \end{array}
\right).
\end{align}
Similarly we can write matrix forms for other operators.  
Now we can calculate $\Gamma(\hat{S}_1^\textsc{a},\hat{S}_1^\textsc{b})$ using the density matrix \eqref{rhoMatrixFormSbasis} and the matrix forms of the appropriate $\hat{S}^\nu_m$. 
For example, the correlation $\Gamma(\hat{S}^\textsc{b}_1,\hat{S}^\textsc{a}_1)$ is given by:
\begin{equation}
\label{CorrelationX-X}
  \Gamma(\hat{S}_1^\textsc{a},\hat{S}_1^\textsc{b}) = \Tr[\rho_\textsc{ab}S_1^\textsc{a}S_1^\textsc{b}] - \Tr[\rho_\textsc{ab} S_1^\textsc{b}]\Tr[\rho_\textsc{ab}S_1^\textsc{a}] = \frac{1}{2}(\Re[\rho_{23}]+\Re[\rho_{14}]) - \frac{1}{4}(\Re[\rho_{13}]+ \Re[\rho_{24}])(\Re[\rho_{12}]+\Re[\rho_{34}]).
\end{equation}
We calculate the other correlations in a similar fashion, and the expression for them in terms of the elements of the density matrix $\rho_{ab}$ are given below:
\begin{align}
\label{CorrelationOb-Ob}
    & \Gamma(\hat{S}_1^\textsc{a},\hat{S}_2^\textsc{b}) = \frac{1}{2}(-\Im[\rho_{14}]+\Im[\rho_{23}]) - \frac{1}{4}(\Re[\rho_{13}]+\Re[\rho_{24}])(-\Im[\rho_{12}]-\Im[\rho_{34}]), \\
    & \Gamma(\hat{S}_1^\textsc{a},\hat{S}_3^\textsc{b}) = \frac{1}{2}(\Re[\rho_{13}]-\Re[\rho_{24}]) \\
    & \Gamma(\hat{S}_2^\textsc{a},\hat{S}_1^\textsc{b}) =  -\frac{1}{2}(\Im[\rho_{23}]+\Im[\rho_{14}]) + \frac{1}{4}(\Im[\rho_{13}]+\Im[\rho_{24}])(\Re[\rho_{12}]+\Re[\rho_{34}]), \\
    & \Gamma(\hat{S}_2^\textsc{a},\hat{S}_2^\textsc{b}) = \frac{1}{2}(\Re[\rho_{23}]-\Re[\rho_{14}]) + \frac{1}{4}(\Im[\rho_{13}]+\Im[\rho_{24}])(-\Im[\rho_{12}]-\Im[\rho_{34}]), \\
    & \Gamma(\hat{S}_2^\textsc{a},\hat{S}_3^\textsc{b}) = \frac{1}{2}(-\Im[\rho_{13}]+\Im[\rho_{24}]) , \\
    & \Gamma(\hat{S}_3^\textsc{a},\hat{S}_1^\textsc{b}) =
    \frac{1}{2}(\Re[\rho_{12}]-\Re[\rho_{34}])\\
    & \Gamma(\hat{S}_3^\textsc{a},\hat{S}_2^\textsc{b}) =
    \frac{1}{2}(-\Im[\rho_{12}]+\Im[\rho_{34}])\\
    &\Gamma(\hat{S}_3^\textsc{a},\hat{S}_3^\textsc{b}) = 0. 
\end{align}
Using the density matrix elements from Eq. \eqref{rho matrix values S basis}, and the expressions for various $\Gamma(\hat{S}_m^\textsc{a},\hat{S}_n^\textsc{b})$ given above, we can evaluate the expression for correlation between any two local operators $\hat{\mathcal{O}}^\textsc{a}$ and $\hat{\mathcal{O}}^\textsc{b}$ in Eq. \eqref{CorrelationOf-Ol} to obtain the general expression for $\Gamma(\hat{\mathcal{O}}^\textsc{a},\hat{\mathcal{O}}^\textsc{b})$, Eq. \eqref{correlation General}.

A particularly interesting correlation to calculate is the correlation between the Hamiltonians of the detectors A and B. 
Note that for any detector $\nu$, its Hamiltonian is given by:
\begin{equation}
    \hat{H}_\nu = \Omega_\nu\ket{e_\nu}\!\bra{e_\nu} = \frac{\Omega_\nu}{2} \hat{\openone}+\frac{-\Omega_\nu}{2} \hat{S}_1^\nu.
\end{equation}
Thus, in Eq. \eqref{correlation General} we use $\zeta =\zeta_\textsc{bb}=\zeta_\textsc{aa}$, $a_1=\frac{-\Omega_\textsc{a}}{2}$, $b_1=\frac{-\Omega_\textsc{b}}{2}$ and $a_2=a_3=b_2=b_3=0$, to obtain,

\begin{equation}
\label{Correlation H-H}
    \Gamma(\hat{H}_\textsc{a},\hat{H}_\textsc{b}) = \Big(\frac{\Omega_\textsc{b}}{2}\Big)\Big(\frac{\Omega_\textsc{a}}{2}\Big)\Big(\prod_{j}\cos{\xi_{j\textsc{b}}}\Big)e^{-\zeta} \Big[(\cosh{\zeta_\textsc{ab}}-\cos{\xi_\textsc{ab}})\cos{\xi_{0\textsc{b}}}\cos{\xi_{0\textsc{a}}}+\sinh{\zeta_\textsc{ab}}\sin{\xi_{0\textsc{b}}}\sin{\xi_{0\textsc{A}}}  \Big].
\end{equation}

\section{Calculating mutual information and quantum discord}
\label{appendix: X-state}

The calculation of quantum discord \eqref{DiscordDef} involves the calculation of $\mathcal{C}$-correlation function and mutual information.
The calculation of mutual information \eqref{mutualInfoDef} turns out to be relatively simpler as it only depends on the eigenvalues of the joint density matrix and the density matrices of reduced subsystems.
For $\mathcal{C}$-correlation function \eqref{ClassicalCorrDef} however, analytical expressions occur only in a few cases such as for $X$-state density matrices \cite{ali2010quantum}, as the optimization over the set of von Neumann measurements is harder in general.
When the initial state of the field is the vacuum state (i.e $T_{0\nu}$ =0), the density matrix of target detectors indeed turns out to be an $X$-state in the basis $\{\ket{g_\textsc{a}}\ket{g_\textsc{b}},e^{i\Omega_\textsc{b} t_\textsc{b}}\ket{g_\textsc{a}}\ket{e_\textsc{b}},e^{i\Omega_\textsc{a} t_\textsc{a}}\ket{e_\textsc{a}}\ket{g_\textsc{b}},e^{i\Omega_\textsc{a} t_\textsc{a}}e^{i\Omega_\textsc{b} t_\textsc{b}}\ket{e_\textsc{a}}\ket{e_\textsc{b}} \}$.
we write the matrix form $\sigma_{\tab}$ of the state $\hat{\rho}_{{\textsc{ab}}}$ \eqref{rho fl} in this basis as:
\begin{equation}
\label{rhoMatrixGbasis}
  \sigma_{\textsc{ab}}= \left(
\begin{array}{cccc}
  \sigma_{11} & 0 & 0 &{\sigma}_{14}\\
 0 & \sigma_{22} & \sigma_{23} & 0 \\
 0 & {\sigma}_{23}^* & \sigma_{33} & 0 \\
 \sigma_{14}^* & 0 & 0 & \sigma_{44} \\
\end{array}
\right)  
\end{equation}{}
Defining $\zeta \coloneqq \zeta_\textsc{bb}= \zeta_\textsc{aa}$, the matrix elements are:

\begin{align}
    &\nonumber \sigma_{11} = \frac{1}{4}\big[1 + \mathrm{e}^{-\frac{\zeta}{2}}+ \mathrm{e}^{-\frac{\zeta}{2}}\cos{\xi_{\textsc{ab}}}\prod_{j=1}^N\cos (\xi_{j\textsc{b}}) +  \mathrm{e}^{-\zeta}\cosh{\zeta_\textsc{ab}}\prod_{j=1}^N\cos (\xi_{j\textsc{b}}) \big], \\
   & \nonumber\sigma_{22} = \frac{1}{4}\big[1 + \mathrm{e}^{-\frac{\zeta}{2}}- \mathrm{e}^{-\frac{\zeta}{2}}\cos{\xi_{\textsc{ab}}}\prod_{j=1}^N\cos (\xi_{j\textsc{b}}) -  \mathrm{e}^{-\zeta}\cosh{\zeta_\textsc{ab}}\prod_{j=1}^N\cos (\xi_{j\textsc{b}}) \big], \\
   &\nonumber\sigma_{33} = \frac{1}{4}\big[1 - \mathrm{e}^{-\frac{\zeta}{2}}+ \mathrm{e}^{-\frac{\zeta}{2}}\cos{\xi_{\textsc{ab}}}\prod_{j=1}^N\cos (\xi_{j\textsc{b}}) -  \mathrm{e}^{-\zeta}\cosh{\zeta_\textsc{ab}}\prod_{j=1}^N\cos (\xi_{j\textsc{b}}) \big], \\
   & \nonumber\sigma_{44} = \frac{1}{4}\big[1 - \mathrm{e}^{-\frac{\zeta}{2}}- \mathrm{e}^{-\frac{\zeta}{2}}\cos{\xi_{\textsc{ab}}}\prod_{j=1}^N\cos (\xi_{j\textsc{b}}) +  \mathrm{e}^{-\zeta}\cosh{\zeta_\textsc{ab}}\prod_{j=1}^N\cos (\xi_{j\textsc{b}}) \big], \\
   & \nonumber \sigma_{14} = -\frac{1}{4} \mathrm{e}^{-\zeta} [\ii \mathrm{e}^{\frac{\zeta}{2}}  \sin{\xi_{\textsc{ab}}}  + 
   \sinh{\zeta_{\textsc{ab}}}]\prod_{j=1}^N\cos (\xi_{j\textsc{b}}) , \\
   & \sigma_{23} = \frac{1}{4} \mathrm{e}^{-\zeta} [\ii \mathrm{e}^{\frac{\zeta}{2}} \sin{\xi_{\textsc{ab}}}  +\sinh{\zeta_{\textsc{ab}}}]\prod_{j=1}^N\cos (\xi_{j\textsc{b}}) \label{rhoMatrixGbasisValues}.
\end{align}
We show here the calculation of quantum discord and mutual information subsequently.

\subsection{Mutual information}\label{appendix:mutualInfo}
Recall that the definition of mutual information is given by Eq. \eqref{mutualInfoDef}. 
To calculate it, we use the matrix form $\sigma_{ab}$ of the state $\hat{\rho}_{{\textsc{ab}}}$
It is given by 
 \begin{equation}
     \mathcal{I}(\sigma_{\textsc{ab}}) = S(\sigma_\textsc{a})+S(\sigma_\textsc{b}) + \sum_{j=0}^3 \lambda_j \log_2 \lambda_j.
 \end{equation}

Here $\lambda_j$ are the eigenvalues of the matrix $\sigma_{\textsc{ab}}$
We first compute $S(\sigma_\textsc{a})$ and $S(\sigma_\textsc{b})$, the entropies of each of the reduced states: 

\begin{align}
    S(\sigma_\textsc{a}) &= - [(\sigma_{11}+ \sigma_{22})\log_2(\sigma_{11}+ \sigma_{22}) + (\sigma_{33}+ \sigma_{44})\log_2(\sigma_{33}+ \sigma_{44}) ], \\
    S(\sigma_\textsc{b}) &= - [(\sigma_{11}+ \sigma_{33})\log_2(\sigma_{11}+ \sigma_{33}) + (\sigma_{22}+ \sigma_{44})\log_2(\sigma_{22}+ \sigma_{44}) ].
\end{align}
It's straight forward to obtain: 
\begin{align}
    \sigma_{11}+ \sigma_{33} = \frac{1}{2}\bigg(1 + \mathrm{e}^{-\frac{\zeta}{2}}\cos{\xi_\textsc{ab}}\prod_{j=1}^N\cos (\xi_{j\textsc{b}})\bigg), \quad \quad  \sigma_{11}+ \sigma_{22} = \frac{1}{2}(1 + \mathrm{e}^{-\frac{\zeta}{2}}), \\
     \sigma_{22}+ \sigma_{44} = \frac{1}{2}\bigg(1 - \mathrm{e}^{-\frac{\zeta}{2}}\cos{\xi_\textsc{ab}}\prod_{j=1}^N\cos (\xi_{j\textsc{b}})\bigg), \quad \quad  \sigma_{33}+ \sigma_{44} = \frac{1}{2}(1 - \mathrm{e}^{-\frac{\zeta}{2}}).
\end{align}
Using the following definition of $h(x)$ \eqref{h(x)}
we obtain 
\begin{align}
\label{SrholAp}
    S(\sigma_\textsc{b}) = h\bigg(\mathrm{e}^{-\frac{\zeta}{2}}\cos{\xi_\textsc{ab}}\prod_{j=1}^N\cos (\xi_{j\textsc{b}})\bigg), \quad \quad
    S(\sigma_\textsc{a}) = g(\mathrm{e}^{-\frac{\zeta}{2}}).
\end{align}

We calculate now the term $\sum_{j=0}^3 \lambda_j \log_2 \lambda_j$. From \cite{ali2010quantum} we note that
\begin{align}
    &\lambda_0 = \frac{1}{2}(a_1 + a_2), \quad  \lambda_1 = \frac{1}{2}(a_1 - a_2), \quad \lambda_2 = \frac{1}{2}(b_1 + b_2), \quad \lambda_3 = \frac{1}{2}(b_1 - b_2), \\
    &a_1 = \sigma_{11}+ \sigma_{44} = \frac{1}{2}\bigg(1+\mathrm{e}^{-\zeta}\cosh \zeta_\textsc{ab}\prod_{j=1}^N\cos (\xi_{j\textsc{b}})\bigg),\\
    &a_2 = \sqrt{(\sigma_{11}- \sigma_{44})^2+ 4 \abs{\sigma_{14}}^2} = \frac{\mathrm{e}^{-\zeta/2}}{2}\sqrt{1 + 2\cos \xi_\textsc{ab}\prod_{j=1}^N\cos (\xi_{j\textsc{b}}) + \Big(1 +e^{-\zeta}\sinh^2\zeta_\textsc{ab}\Big)\bigg(\prod_{j=1}^N\cos (\xi_{j\textsc{b}})\bigg)^2},\\
    &b_1 = \sigma_{22}+ \sigma_{33} = \frac{1}{2}\bigg(1-\mathrm{e}^{-\zeta}\cosh \zeta_\textsc{ab}\prod_{j=1}^N\cos (\xi_{j\textsc{b}})\bigg),\\
    &b_2 = \sqrt{(\sigma_{22}- \sigma_{33})^2+ 4 \abs{\sigma_{23}}^2} =  \frac{\mathrm{e}^{-\zeta/2}}{2}\sqrt{1 - 2\cos \xi_\textsc{ab}\prod_{j=1}^N\cos (\xi_{j\textsc{b}}) +\Big(1 +e^{-\zeta}\sinh^2\zeta_\textsc{ab}\Big)\bigg(\prod_{j=1}^N\cos (\xi_{j\textsc{b}})\bigg)^2}.\\
\end{align}
The term $\sum_{j=0}^3 \lambda_j \log_2 \lambda_j$ can be manipulated to the form it has in equation \eqref{h(x)}. For example the first two terms in the sum can be written as 
\begin{equation}
    \begin{split}
     \lambda_0 \log_2 \lambda_0 + \lambda_1 \log_2 \lambda_1 &= \frac{a_1+a_2}{2}\log_2\Big(\frac{a_1+a_2}{2}\Big) + \frac{a_1-a_2}{2}\log_2\Big(\frac{a_1-a_2}{2}\Big) \\
     & = a_1\Big(\frac{1+a_2/a_1}{2}\log_2\Big(\frac{a_1+a_2}{2}\big) + \frac{1-a_2/a_1}{2}\log_2\Big(\frac{a_1-a_2}{2}\Big)\Big) =  \\
     &= a_1 \log_2(a_1) + a_1\Big(\frac{1+a_2/a_1}{2}\log_2\Big(\frac{1+a_2/a_1}{2}\Big) +\frac{1-a_2/a_1}{2}\log_2\Big(\frac{1-a_2/a_1}{2}\Big)\Big)\\
     & = a_1 \log_2(a_1) - a_1 h(a_2/a_1).
    \end{split}
\end{equation}
Similarly, $ \lambda_2 \log_2 \lambda_2 + \lambda_3 \log_2 \lambda_3 = b_1 \log_2(b_1) - b_1 h(b_2/b_1)$. Therefore we obtain 
\begin{equation}
\label{sumlambdajlog}
   -S(\hat{\rho}_{{\textsc{ab}}})= \sum_{j=0}^3 \lambda_j \log_2 \lambda_j = a_1 \log_2(a_1) +b_1 \log_2(b_1) - a_1 h(a_2/a_1)- b_1 h(b_2/b_1).
\end{equation}
Hence from Eqs. \eqref{SrholAp} and \eqref{sumlambdajlog} we retrieve Eqs. \eqref{h(x)}.

 \subsection{Henderson-Vedral $\mathcal{C}$ correlation function}
 \label{appendix::classCorr}
Here we follow the approach in \cite{ali2010quantum} to calculate the correlation function $\mathcal{C}(\hat{\rho})$.
Recall that it is defined \eqref{ClassicalCorrDef} as 
 \begin{equation}
         \mathcal{C}(\hat{\rho}_{{\textsc{ab}}})= S(\hat{\rho}_\textsc{a}) - \inf_{\{\hat M_k\}}S(\hat{\rho}_{{\textsc{ab}}}| \{\hat M_k\}) =  S(\hat \rho_\textsc{a}) - \inf_{\{\hat M_k\}}\sum_{k\in\{0,1\}}p_k S(\hat{\rho}_{{\textsc{ab}}}^{(k)}).
 \end{equation}
Here $M_k$ are von Neumann measurements on the B subsystem; $M_0 = \hat{V}\ket{g_\textsc{b}}\bra{g_\textsc{b}}\hat{V}^\dagger$ and  $M_1 = \hat{V}\ket{e_\textsc{b}}\bra{e_\textsc{b}}\hat{V}^\dagger$ where $\hat{V} \in SU(2)$ \begin{equation}
     \hat{\rho}_{{\textsc{ab}}}^{(k)} = \frac{1}{p_k}(\openone\otimes \hat M_k) \hat{\rho}_{{\textsc{ab}}} (\openone\otimes \hat M_k), \quad \quad p_k = \Tr[(\openone\otimes \hat M_k) \hat{\rho}_{{\textsc{ab}}} (\openone\otimes M_k)].
 \end{equation}
 Each $\hat V \in SU(2)$ can be written as $  \hat{V} = t\openone+\ii\vec{y}.\vec{\sigma}$ , where $t,y_1,y_2,y_3 \in \mathbb{R}$ satisfy $t^2+y_1^2+y_2^2+y_3^2 =1 $ and $\vec{\sigma}$ denotes the triad of Pauli operators $(\sigma_x,\sigma_y,\sigma_z)$.
 As there is a one-one relation between $SU(2)$ and the set of von Neumann measurements $\{\hat{M}_k\}$, the latter can be characterised by the 3 parameter set of \textit{SU(2)} operators.
 Thus we can simply minimise $S(\hat{\rho}_{{\textsc{ab}}}| \{\hat M_k\})$ over the possible range of parameters $t,y_1,y_2,y_3$.
 To perform our calculations we use the matrix $\sigma_\textsc{ab}$ and all the results are in terms of its matrix elements. 
 We define the parameters $k = t^2 +y_3^2, m=(ty_1+ y_2y_3)^2, n = (ty_2-y_1y_3)(ty_1+y_2y_3 )$ and $l = y_1^2+y_2^2 = 1-k$, which yields a simpler relation for $S(\hat{\rho}_{{\textsc{ab}}}| \{\hat M_k\})$. 
 The probability of an outcome $\hat{\rho}_{\textsc{ab}}^{(k)}$ is given by 
\begin{equation}
 \begin{split}
  & p_0 = [ (\sigma_{11} +\sigma_{33})k+(\sigma_{22} +\sigma_{44})l], \\
  & p_1 = [ (\sigma_{11} +\sigma_{33})l+(\sigma_{22} +\sigma_{44})k]. 
 \end{split}
\end{equation}
The von Neumann entropies of the measurement outcomes $\sigma_\textsc{ab}^{(i)}$ are given by 
\begin{equation}
  S(\sigma_\textsc{ab}^{(0)}) = h(\theta), \quad \quad  S(\sigma_\textsc{ab}^{(1)}) = h(\theta').
 \end{equation}
 Note that, $h(x)$ has been defined in Eq. \eqref{h(x)}. $\theta$ and $\theta'$ in the above equation are given by
 \begin{align}
 & \theta = \sqrt{\frac{[(\sigma_{11}-\sigma_{33})k +(\sigma_{22}-\sigma_{44})l]^2+\Theta}{[(\sigma_{11}+\sigma_{33})k +(\sigma_{22}+\sigma_{44})l]^2}}, \quad \quad \quad \theta' = \sqrt{\frac{[(\sigma_{11}-\sigma_{33})l +(\sigma_{22}-\sigma_{44})k]^2+\Theta}{[(\sigma_{11}+\sigma_{33})l +(\sigma_{22}+\sigma_{44})k]^2}}. 
\end{align}
 where $\Theta = 4kl[\abs{\sigma_{14}}^2 + \abs{\sigma_{23}}^2+ 2\Re(\sigma_{14}\sigma_{23}^*)] -16 m\Re(\sigma_{14}\sigma_{23}^*) + 16 n\Im(\sigma_{14}\sigma_{23}^*)$, according to \cite{erratum}.
 
 According to \cite{ali2010quantum}, the minimum value of $S(\hat{\rho}_{{\textsc{ab}}}|\{\hat{M}_k\})$  occurs in one of the following cases : 1) $k=1,l=0(m=0,n=0) $ or 2) $k=l=\frac{1}{2}, m \in \{ 0, \frac{1}{4} \}, n \in \{ 0, \pm \frac{1}{8}\} $.
 First we look at the case $k=l= 1/2$.
 It is easy to see that for this case, $p_0 =p_1 = 1/2$ and $\theta = \theta' = \sqrt{[\sigma_{11}+\sigma_{22}-\sigma_{33}-\sigma_{44}]^2+ 4\Theta}$. 
 Thus, we need to simply minimise $h(\theta)$ . For positive $\theta, h(\theta)$ is a monotonically decreasing function. Thus we simply must chose the maximum value of $\theta$ out of the 6 possibilities for \textit{m} and \textit{n}.
 Using the elements of the density matrix from \eqref{rhoMatrixGbasisValues}, we find that 

 \begin{equation}
 \theta = e^{-\zeta/2}\sqrt{1 + 4me^{-\zeta}[\sinh^2 \zeta_{\textsc{ab}}+ e^{\zeta}\sin^2\xi_\textsc{ab}]\bigg(\prod_{j=1}^N\cos (\xi_{j\textsc{b}})\bigg)^2}.
 \end{equation}
Clearly, theta increases with $m$, so $m=1/4$ is the appropriate choice.
 Thus, when $k=l=1/2$ we have
 \begin{equation}
 \label{CCorMinimaOpt1}
 S(\hat{\rho}_{{\textsc{ab}}}| \{\hat M_k\}) = g\Bigg(e^{-\zeta/2}\sqrt{1 +e^{-\zeta}[\sinh^2 \zeta_{\textsc{ab}}+ e^{\zeta}\sin^2\xi_\textsc{ab}]\bigg(\prod_{j=1}^N\cos (\xi_{j\textsc{b}})\bigg)^2}\:\: \Bigg).
 \end{equation}
Now, we look at the case when $k=1, l=0$. This is same as the case $k=0, l=1$.
In either cases, $m=n=0$.
Thus $\Theta =0$. 
By simply substituting the values of the density matrix elements, we get,

\begin{equation}
\label{CCorMinimaOp2}
\begin{split}
   & S(\hat{\rho}_{{\textsc{ab}}}| \{\hat M_k\})  = p_0h(\theta_0)+p_1h(\theta_1),\\
     & p_0 = \frac{1}{2}(1+e^{-\zeta/2}\cos{\xi_\textsc{ab}}\prod_{j=1}^N\cos (\xi_{j\textsc{b}})),\\
     & p_1 = \frac{1}{2}(1-e^{-\zeta/2}\cos{\xi_\textsc{ab}}\prod_{j=1}^N\cos (\xi_{j\textsc{b}})),\\
     & \theta_0 =\frac{\Bigg|\mathrm{e}^{-\zeta/2}+\mathrm{e}^{-\zeta}\cosh \zeta_\textsc{ab}\prod_{j=1}^N\cos (\xi_{j\textsc{b}})\Bigg|}{2 p_0},\\
     & \theta_1  =\frac{\abs{\mathrm{e}^{-\zeta/2}-\mathrm{e}^{-\zeta}\cosh \zeta_\textsc{ab}\prod_{j=1}^N\cos (\xi_{j\textsc{b}})}}{2 p_1}
\end{split}
\end{equation}
Chosing the minimum out of the two values of $S(\hat{\rho}_{{\textsc{ab}}}| \{\hat M_k\})$ from Eq. \eqref{CCorMinimaOpt1} and \eqref{CCorMinimaOp2}, we get the formula for the correlation function $\mathcal{C}(\rhoh)$ mentioned in Eq. \eqref{CCorrFinal}.

\subsection{Optimality of discord formula}
\label{discordCheckFormula}

As mentioned before in section \ref{subsec-Discord}, to check if the above algorithm to calculate $\mathcal{C}(\rhoh)$, and subsequently quantum discord, is correct we need to prove that one of two conditions \eqref{condition1}, \eqref{condition2} for a two-qubit system  is correct ~\cite{Chen_2011}. To check those conditions we first have to transform our density matrix \eqref{rhoMatrixGbasis} to a real matrix via a local unitary operation. 
One unitary that succeeds in this transformation is $U_\textsc{A}\otimes \openone_\textsc{b}$, with
\begin{equation}
    U_\textsc{a} = \left(
\begin{array}{cc}
  u_{00} & 0 \\
  0 & u_{11} \\
\end{array}
\right).
\end{equation}
The only condition needed is that $\abs{u_{00}}=\abs{u_{11}}=1$ and that $\Im(u_{00}u_{11}^*\sigma_{23}) = 0$. Therefore one possible choice is $u_{11}=1$ and
\begin{equation}
    u_{00} =  \frac{ \sinh{\zeta_{\textsc{ab}}}-\ii \mathrm{e}^{\frac{\zeta}{2}} \sin{\xi_{\textsc{ab}}} }{\sqrt{ \sinh^2{\zeta_{\textsc{ab}}}+ \mathrm{e}^{\zeta} \sin^2{\xi_{\textsc{ab}}} }}
\end{equation}
In that case, the density matrix \eqref{rhoMatrixGbasis} transforms to: 
\begin{equation}
  \sigma_{\textsc{ab}}= \left(
\begin{array}{cccc}
  \sigma_{11} & 0 & 0 &-\abs{\sigma_{23}}\\
 0 & \sigma_{22} & \abs{\sigma_{23}} & 0 \\
 0 & \abs{\sigma_{23}} & \sigma_{33} & 0 \\
-\abs{\sigma_{23}} & 0 & 0 & \sigma_{44} \\
\end{array}
\right)  
\end{equation}{}
with the same matrix entries as listen in \eqref{rhoMatrixGbasisValues}.
Following \cite{Chen_2011} the correlation function $\mathcal{C}(\rhoh)$ calculated in \eqref{CCorrFinal} is correct if one of the two following conditions, corresponding to $\sigma_z^\textsc{a}$ or $\sigma_x^\textsc{a}$ being the optimal measurement respectively are satisfied:
\begin{align}
    &4(\abs{\sigma_{23}}^2)\leq (\sigma_{11}-\sigma_{22})(\sigma_{44}-\sigma_{33}),\\
    &\abs{\sqrt{\sigma_{11}\sigma_{44}}-\sqrt{\sigma_{22}\sigma_{33}}}\leq 2\abs{\sigma_{23}}.
\end{align}
Using the matrix elements from equations \eqref{rhoMatrixGbasisValues} we see
\begin{equation}
    4(\abs{\sigma_{23}}^2- (\sigma_{11}-\sigma_{22})(\sigma_{44}-\sigma_{33}) = \frac{e^{-2\zeta}}{4}(e^{\zeta}-1)\prod_{j=1}^N\cos^2 (\xi_{j\textsc{b}}) \geq 0,
\end{equation}
as $\zeta$ is always positive. Thus the first condition is always false, independent of parameters.
The second condition however is met for all instances where we calculate the quantum discord (i.e figures \ref{XTdepzoom}, \ref{XTdepzoomGauss}, \ref{strengthdep}).
To show that, we plot the functional $f[{\sigma_\tab}] =\abs{\sqrt{\sigma_{11}\sigma_{44}}-\sqrt{\sigma_{22}\sigma_{33}}}- 2\abs{\sigma_{23}}$ for the different parameters we consider for obtaining figures \ref{XTdepzoom}, \ref{XTdepzoomGauss}, \ref{strengthdep}, and find that  $f[{\sigma_\tab}]$ is negative for all these cases. The results are shown in figures \ref{checkDiscordXT}, \ref{checkDiscordStrength}.

\begin{figure*}
\begin{tabular}{cc}
    \includegraphics[width=0.4\textwidth]{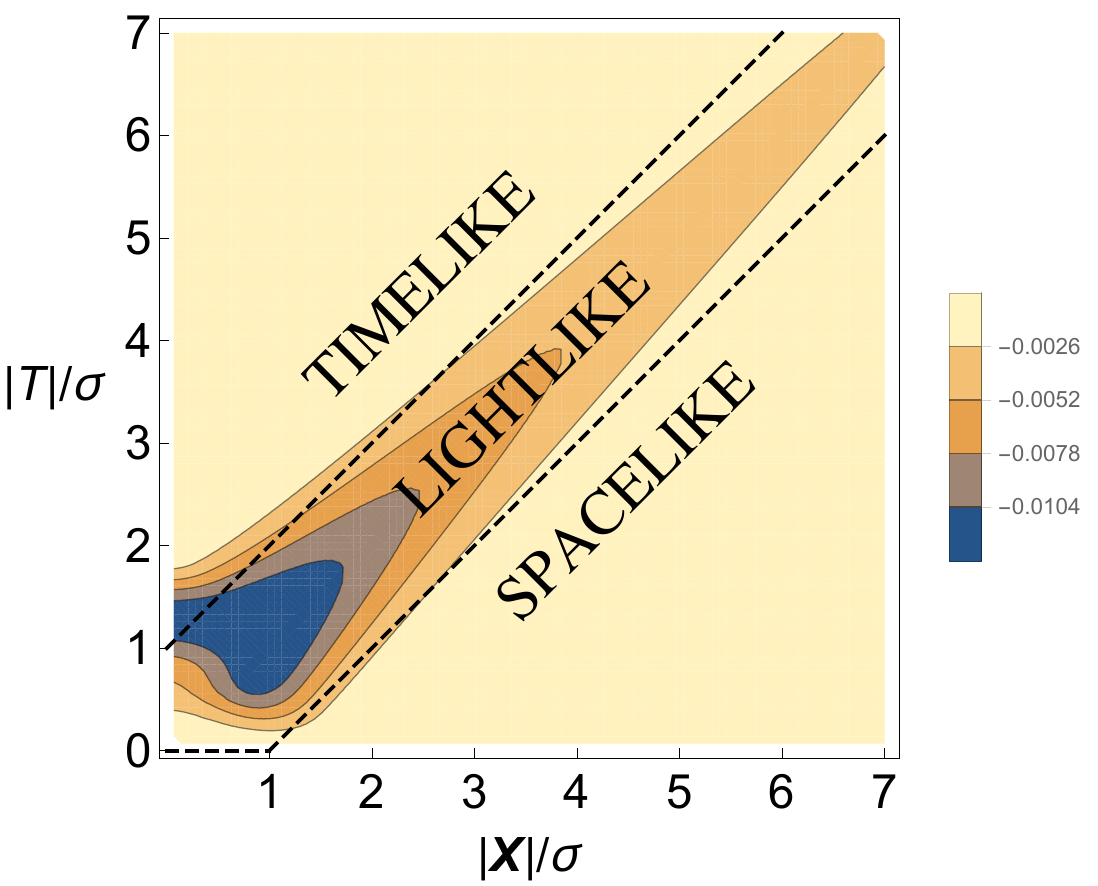}
    &
    \includegraphics[width=0.4\textwidth]{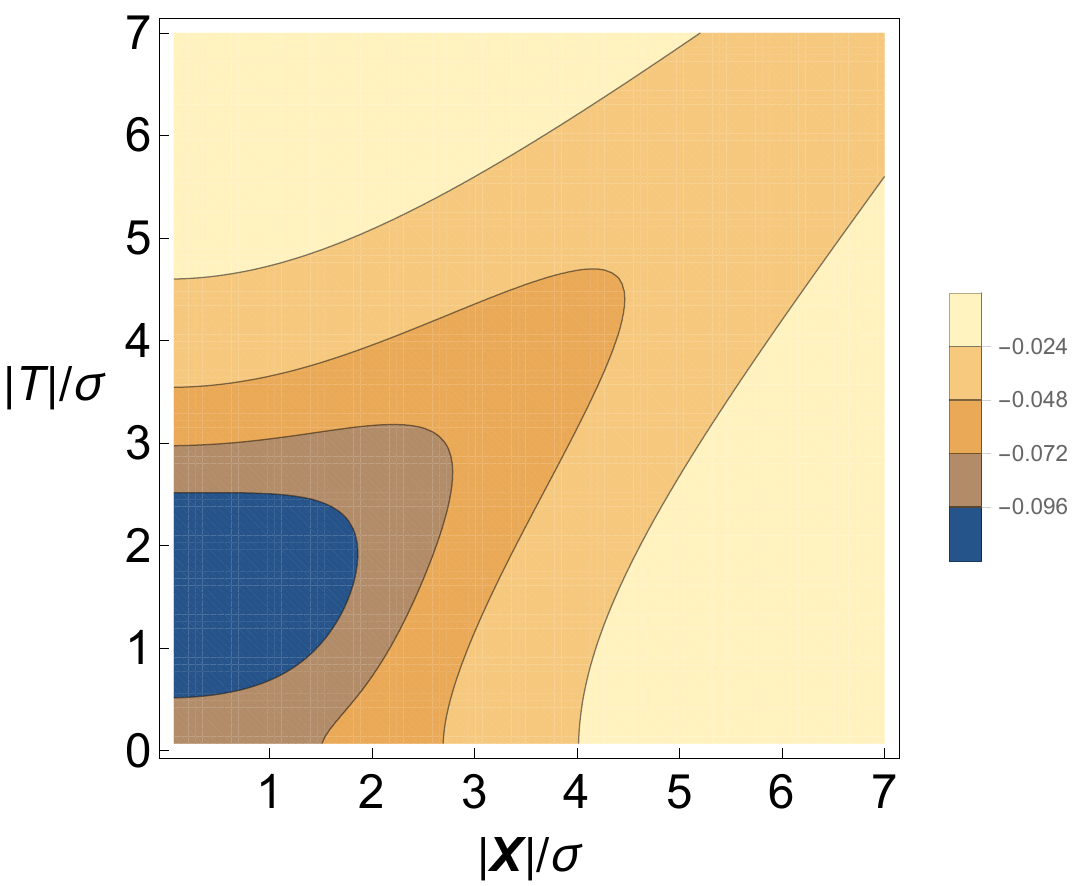}
  \\
(a) Hard-sphere smearing & (b) Gaussian smearing 
\end{tabular}
\caption{Plot showing variation of  $f[{\sigma_\tab}]$ with \mbox{$\abs{\bm{X}} = \abs{\bm{x}_\textsc{b} - \bm{x}_\textsc{a}}$, and $T = t_\textsc{b} - t_\textsc{a}$}. In (a) we consider a Hard-sphere smearing function for the detectors with the same parameters as those used to obtain figure \ref{XTdepzoom}, and in (b) we consider a Gaussian smearing function for the detectors with the same parameters as those used to obtain figure \ref{XTdepzoomGauss}. $f[{\sigma_\tab}]$ being negative in both cases validates the use of equation  \eqref{CCorrFinal} to compute $\mathcal{C}(\rhoh)$ and quantum discord for figures \ref{XTdepzoom} and \ref{XTdepzoomGauss} }
\label{checkDiscordXT}
\end{figure*}

\begin{figure*}
  \includegraphics[width=0.6\textwidth]{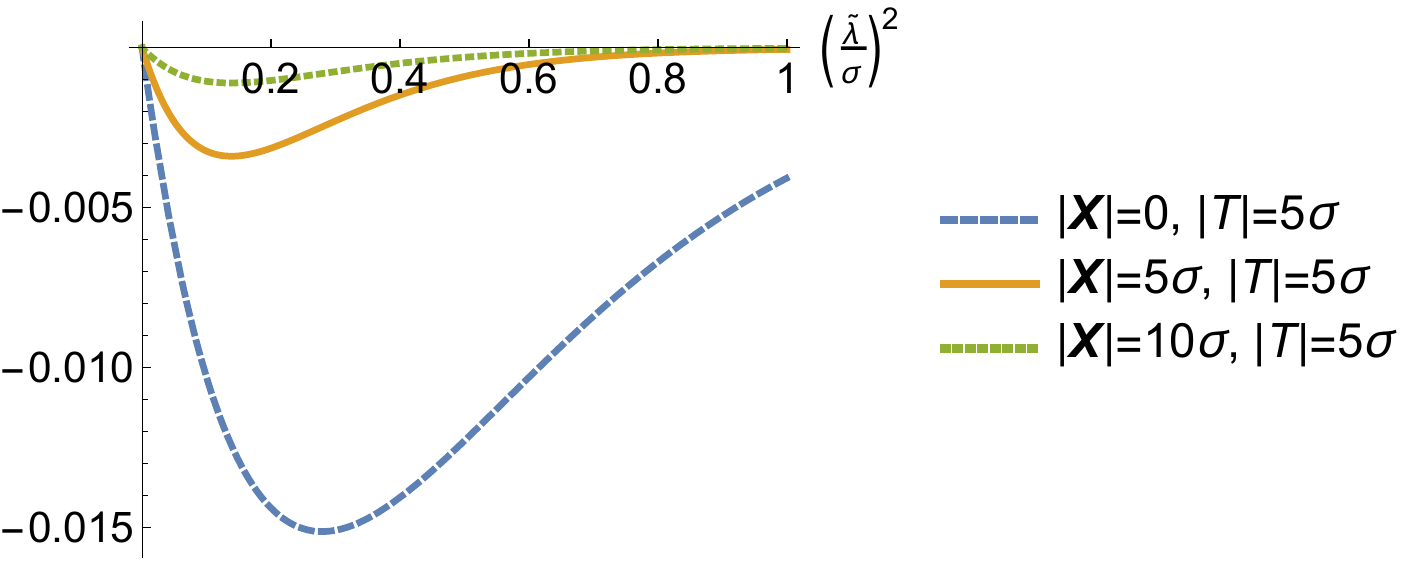}
\caption{Plot showing variation of  $f[{\sigma_\tab}]$ with $\tilde\lambda_\textsc{a}\tilde\lambda_\textsc{b}$, the product of the effective coupling strengths.
We consider a Hard-sphere smearing function for the detectors with the same parameters as those used to obtain figure \ref{strengthdep}. $f[{\sigma_\tab}]$ being negative validates the use of equation  \eqref{CCorrFinal} to compute $\mathcal{C}(\rhoh)$ and quantum discord for figure \ref{strengthdep}.}
\label{checkDiscordStrength}
\end{figure*}

\section{Calculation of $T_{ij}$}
\label{appendix:Tab}
Here we intend to evaluate the integral in the definition of $T_{ij}$ in \eqref{Tij}. Recall, that 
\begin{equation}
    \begin{split}
        & \beta_\nu(\bm{k})= -\ii \tilde\lambda_\nu\frac{\Tilde{F_\nu}(-\bm{k})}{\sqrt{2|\bm{k}|}}e^{\ii(|\bm{k}|t_\nu - \bm{k}\cdot\bm{x_\nu})},\\
        &  T_{ij}= \frac{\zeta_{ij}}{4}+ \ii\frac{\xi_{ij}}{4} := \int \dd^n\bm{k}\beta^*_i(\bm{k})\beta_j(\bm{k}).
    \end{split}
\end{equation}
Plugging $\beta_\nu(\bm{k})$ back in the expression for $T_{ij}$, we see that the required integral can be cast in the following form:
\begin{equation}
\label{Tab}
    T_{ij} = \tilde{\lambda}_i\tilde{\lambda}_j\int \dd^n\bm{k} \frac{\tilde{F}_i(\bm{k})\tilde{F}_j(-\bm{k})}{2|\bm{k}|}e^{\ii(|\bm{k}|T-\bm{k}\cdot\bm{X})},
\end{equation}
where $\bm{X}=\bm{x_j}-\bm{x_i}$ and $T=t_j-t_i$. 
When $i=j$,  $\bm{X}=0$ and $T=0$ and thus $T_{ij}$ is real, and independent of the spcetime locations of detectors $i$ and $j$.
Particularly when the detectors are considered to be identical, as in \ref{sectionHarvestingCorrelations}, the quantity $\zeta_{ii} \coloneqq 4T_{ii}$. 
is the same for every detector.
Thus we omit the subscripts and simply call it $\zeta$.

The product of the two Fourier transforms $\tilde{F}_i(\bm{k})\tilde{F}_j(-\bm{k})$  can be written as the Fourier transform of a convolution (we are assuming that $\tilde{F}$ is a real function), in the following way:
\begin{equation}
 \tilde{F}_i(\bm{k})\tilde{F}_j(-\bm{k}) = \frac{1}{\sqrt{(2\pi)^n}}\int\int \; \dd^n\bm{x'}\dd^n\bm{z'}F_j(\bm{z'})F_i(\bm{x'}+\bm{z'})e^{\ii\bm{k}\cdot\bm{x'}}.
\end{equation}
Plugging it into equation \eqref{Tab} we get the expression
\begin{equation}
\label{Tab1}
    T_{ij} =  \frac{\tilde{\lambda}_i\tilde{\lambda}_j}{\sqrt{(2\pi)^n}}\int\int \dd^n\bm{x'}\dd^n\bm{z'}\;F_j(\bm{z'})F_i(\bm{x'}+\bm{z'})\int \dd^n\bm{k}\frac{e^{\ii|\bm{k}|T+\ii\bm{k}\cdot(\bm{x'}-\bm{X})}}{2\abs{\bm{k}}}.
\end{equation}
    The integral over $\bm{k}$ can be partially simplified using the properties of Bessel functions of the first kind, $J_n(x)$, 
\begin{equation}
\label{Kintegral}
    \int \dd^n\bm{k}\frac{e^{\ii|\bm{k}|T+\ii\bm{k}\cdot(\bm{x'}-\bm{X})}}{2\abs{\bm{k}}} =\frac{\sqrt{(2\pi)^n}}{2} \int_0^{\infty} \dd k\; \bigg(\frac{k}{\abs{\bm{x'-\bm{X}}}}\bigg)^{\frac{n}{2}-1}J_{\frac{n}{2}-1}\bigg(k\abs{\bm{x'-\bm{X}}}\bigg)e^{\ii kT},
\end{equation}
where $k=\abs{\bm{k}}$. We perform the integral in \ref{Kintegral} for the specific case of 3 space dimensions ($n=3$), in which case the required Bessel function is $J_{1/2}(x)=\sqrt{\frac{2}{\pi x}}\sin x$.
Thus for $n=3$, Eq. \eqref{Kintegral} becomes
\begin{equation}
    \int \dd^3\bm{k}\frac{e^{\ii|\bm{k}|T+\ii\bm{k}\cdot(\bm{x'}-\bm{X})}}{2\abs{\bm{k}}}
    = \frac{2\pi}{\abs{\bm{x'}-\bm{X}}}\int_0^\infty \dd k \; \sin{(k\abs{\bm{x'}-\bm{X}})}e^{\ii kT}.
\end{equation}
To calculate the above, we make use of the following identities:
\begin{equation}
  \begin{split}
        & e^{\ii k\alpha} = \cos{k\alpha}+\ii\sin{k\alpha},\\
        & \int_0^\infty \dd k \cos{k\alpha} = \pi\delta(\alpha), \\
        & \int_0^\infty \dd k \sin{k\alpha} =
        \left\{
        \begin{array}{ll}
            \frac{1}{\alpha} & \quad \alpha \neq 0 \\
            0 & \quad \alpha = 0
        \end{array}
        \right\}.
  \end{split}
\end{equation}
Now, we can write $\sin{(k\abs{\bm{x'}-\bm{X}})}$ using $e^{\ii k\abs{\bm{x'}-\bm{X}}}$ and using the above identities, we find out that: 
\begin{equation}
\label{Kintn=3}
   \PV \int \dd^3\bm{k}\frac{e^{\ii|\bm{k}|T+\ii\bm{k}\cdot(\bm{x'}-\bm{X})}}{2\abs{\bm{k}}}
    = \frac{2\pi}{\abs{\bm{x'}-\bm{X}}}\Bigg[\frac{\abs{\bm{x'}-\bm{X}}}{\abs{\bm{x'}-\bm{X}}^2-T^2}+\ii \frac{\pi}{2}\Big[\delta\bigg(\abs{\bm{x'}-\bm{X}}-T\bigg)-\delta\bigg(\abs{\bm{x'}-\bm{X}}+T\bigg)\Big]\Bigg].
\end{equation}
Note that since $\abs{\bm{x'}-\bm{X}}$ is always non-negative, hence only one of the Dirac deltas in the imaginary part will contribute to $T_{ij}$, depending upon the sign of T.
We have thus performed the $k$ integral for a 3-D space. This integration does not depend on the choice of smearing functions of the detector. 
From this point on, we need to consider specific examples of smearing functions to compute $T_{ij}$ using Eqs. \eqref{Tab1} and \eqref{Kintn=3}.

In Eq. \eqref{Kintn=3},   we see that if  two detectors 1 and 2 are spacelike separated, the delta functions in the imaginary part evaluate to zero and thus $\xi_{12}$ is identically zero. Let us remember that the
Unruh-DeWitt Hamiltonian between the field and each of the detectors was given by
\begin{equation}
\label{unruhAgain}
    \hat{H}(t) = \sum_{i=1}^N {\lambda}_i \chi_i(t) \hat{\mu}_i(t)\otimes \int \dd^n\bm{x} F_i (\bm{x}-\bm{x_i}) \hat{\phi}(t,\bm{x}).
\end{equation}
In order for two detectors to be completely spacelike separated their smearing and switching functions should be compactly supported. Let the support of the smearing function of detector $i$ be $\mathcal{D}_i$. Then $F_i (\bm{x})\neq 0$ iff  $\bm{x}\in \mathcal{D}_i$. From Eq.~\eqref{unruhAgain}, we see this means that the detector couples to the field at the points $\bm{x}-\bm{x_i} \in \mathcal{D}_i$.
We say that two detectors $1$ and $2$ are are spacelike separated if all the spacetime points (events) in $\text{supp}[\chi_1F_1]$ are spacelike separated from all the events in   $\text{supp}[\chi_2F_2]$. Without loss of generality, let us consider a reference frame $(t,\bm x)$ so that detectors $1$ and $2$ instantaneously (delta) couple to the field at times $t_1 < t_2$. Let us call the time interval between switchings $T = t_2-t_1 > 0$. In this coordinate system the points in $\text{supp}[\chi_1F_1]$ are represented by $(t_1, \bm{x_1})$, with $\bm{z_1}-\bm{x_1} \in \mathcal{D}_1$ and any point in spacetime associated to the detector $2$ will be $( t_2,\bm{z_2})$, with $\bm{z_2}-\bm{x_2} \in \mathcal{D}_2$. The spacelike separation condition then reduces to:
\begin{align}
    \abs{\bm{z_1}-\bm{z_2}} > (t_2-t_1) \Rightarrow \abs{\bm{z_1}-\bm{x_1}+\bm{x_1} -\bm{z_2}+\bm{x_2}-\bm{x_2}} > T \Rightarrow \abs{\bm{z_1}-\bm{x_1} -\bm{z_2}+\bm{x_2}-\bm{X}} > T, 
\end{align}
with $\bm{X}=\bm{x_2}-\bm{x_1} $. In light of this, we can now take a convenient change of variables in Eq.~\eqref{Tab1} to understand when $T_{ij}$ is different from zero. Without loss of generality, we can consider $i=1$ and $j=2$ in Eq.~\eqref{Tab1}.  Rewriting $\bm{z}' = \bm{z_2}-\bm{x}_2$ and $\bm{z}' +\bm{x}'= \bm{z_1}-\bm{x}_1$, we observe that the product $F_2(\bm{z}')F_1(\bm{x}'+\bm{z}') = F_2(\bm{z_2}-\bm{x}_2)F_1(\bm{z_1}-\bm{x}_1)$ can only be different from zero when $\bm{z_2}-\bm{x_2}\in \mathcal{D}_2$ and $\bm{z_1}-\bm{x_1}\in \mathcal{D}_1$ simultaneously, which in turns means that  $\bm{z_2}-\bm{x_2}$ and $\bm{z_1}-\bm{x_1}$ are necessarily spacelike separated. The spacelike separation condition provide us with the relation $\bm{x}'$ and $\bm{z}'$ need to meet for $T_{12}$ to be different from zero:
\begin{align}
\label{conditionSpacelike}
    \abs{\bm{z}' +\bm{x}' -\bm{z}'-\bm{X}}=  \abs{\bm{z_1}-\bm{x}_1 -\bm{z_2}+\bm{x}_2-\bm{X}} > T \rightarrow \abs{\bm{x}'-\bm{X}}> T.
\end{align}
If now we look at Eqs.~\eqref{Tab1} and~\eqref{Kintn=3}, we see that the imaginary part of $T_{12}$ will be zero since the arguments of the deltas do not vanish at any spacetime point satisfying condition \eqref{conditionSpacelike}. Therefore we conclude $\xi_{12} = 0$, which in turns means that the geometric factor $\mathcal{I}$ given in the main text is zero when the detectors are spacelike separated.

\subsection{Hard sphere smearing function}
First we consider the smearing function to be a normalised hard sphere for a 3 dimensional case:
\begin{equation}
    F(\bm{x}) =\left\{
        \begin{array}{ll}
            \frac{3}{4\pi\sigma^3} & \quad \abs{\bm{x}} \leq \sigma \\
            0 & \quad \abs{\bm{x}} > \sigma
            
        \end{array}
    \right\},
\end{equation}
where $F(\bm{x})$ is $0$ outside a sphere of radius $\sigma$ centered at the origin.
\subsubsection{Imaginary part}
For the case of 3 spatial dimensions, $\xi_{ij}$ which is 4 times the imaginary part of $T_{ij}$ is calculated from Eqs. \eqref{Kintn=3} and \eqref{Tab1}, and is given by:
\begin{equation}
    \xi_{ij}= \tilde{\lambda}_i\tilde{\lambda}_j\sqrt{2\pi}\int\int \dd^3\bm{x'}\;d^3\vb{z'}\; F_j(\bm{z'})F_i(\bm{x'}+\bm{z'})\frac{T\delta(\abs{\bm{x'}-\bm{X}}-\abs{T})}{\abs{T}\abs{\bm{x'}-\bm{X}}}.
\end{equation}
Plugging this smearing function into the above, we notice that the integral over $\bm{z}$ is simply the volume of intersection between two n-spheres of radius $\sigma$ separated by a distance $x'$. 
Thus the $\bm{z'}$ integral is easily performed, and in particular the for the 3 dimensional case it turns out to be $\frac{3}{8\pi\sigma^3}(1-\frac{x'}{2\sigma})^2(2+\frac{x'}{2\sigma})$ when $x'\leq 2\sigma$ (i.e when there is some intersection between the spheres) and 0 otherwise.
Using this we get
\begin{equation}
    \xi_{ij}= \tilde{\lambda}_i\tilde{\lambda}_j\frac{3\sqrt{2\pi}}{8\pi\sigma^3} \int_{x'\leq 2\sigma} \dd^3\bm{x'} \Big(1-\frac{x'}{2\sigma}\Big)^2\Big(2+\frac{x'}{2\sigma}\Big)\frac{T\delta(\abs{\bm{x'}-\bm{X}}-\abs{T})}{\abs{T}\abs{\bm{x'}-\bm{X}}}.
\end{equation}
To calculate the above we use spherical coordinates and use the variable substitution $\bm{r}=\bm{x'}-\bm{X}$. 
Without loss of generality, we can choose $\bm{X} = (0,0,\abs{\bm{X}})$.  We observe that the original volume of integration is a sphere with radius $2\sigma$ centered in $-\bm{X}$. To get a clearer picture of the integral we can use spherical coordinates, and parametrize $\bm{r}$  as $\bm{r} = r(\cos\phi\sin\theta, \sin\phi\sin\theta, \cos\theta)$ with $\phi\in(0,2\pi)$ and $\theta\in(0, \pi)$. We observe that
\begin{align}
    \abs{\bm{r}+\bm{X}}^2 = r^2+ 2r\abs{\bm{X}}\cos\theta + \abs{\bm{X}}^2.
\end{align}{}
Since we have the term $\delta(r-\abs{T})$, the volume of integration parametrized by $r, \phi, \theta$ will become an area parametrized by $\phi, \theta$. This area will be the intersection of a shell centered in $\bm{0} = (0,0,0)$ with radius $\abs{T}$ (called $S$ henceforth) and the sphere with radius $2\sigma$ centered in $-\bm{X}$ (called $C$ henceforth).
Clearly, the integral will be 0 whenever there is no part of the shell $S$ inside the Sphere $C$. 
Thus we need to carefully evaluate the integral depending upon whether \textit{S} is partially or completely inside \textit{C}. 

Let us first consider the case when $\abs{T}\geq 2\sigma$. 
In picture \ref{fig:my_label} we represent this situation:
\begin{figure}[h]
    \centering
   \includegraphics[scale=0.45]{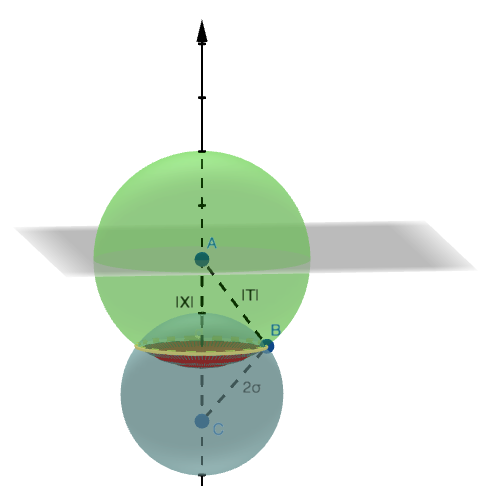}
    \caption{The origin of coordinates is the point A and the axis represented is axis z. We have the original volume of integration, the blue sphere. Since we have the term $\delta(r-\abs{T})$ we see that the integration area will only be the the red one. 
}
    \label{fig:my_label}
\end{figure}
Looking at the picture it is evident that this integral is non-trivial iff $\abs{T}-2\sigma \leq \abs{\bm{x}} \leq \abs{T}+2\sigma$. 
Otherwise, \textit{S} is either completely outside \textit{C} or engulfs \textit{C}.
Either way, there is no part of \textit{S} inside \textit{C}.
However, when this condition is satisfied, there is a partial intersection between \textit{S} and \textit{C}.
The integral, taking the dirac delta, and considering the symmetry in the x-z axis,  thus becomes: 
\begin{equation}
\label{Yab integral}
\begin{split}
    \xi_{ij} = \tilde\lambda_i\tilde\lambda_j\frac{3T\sqrt{2\pi}}{4\sigma^3}  \int_{\theta_0}^{\theta_1 = \pi} \dd\theta \sin{\theta} \Bigg(1-\frac{\sqrt{\abs{T}^2+ 2\abs{T}\abs{\bm{X}}\cos\theta + \abs{\bm{X}}^2}}{2\sigma}\Bigg)^2\Bigg(2+\frac{\sqrt{\abs{T}^2+ 2\abs{T}\abs{\bm{X}}\cos\theta + \abs{\bm{X}}^2}}{2\sigma}\Bigg),
\end{split}   
\end{equation}
where $\cos{\theta_0} = \cos(\pi-\angle{BAC}) = - \cos(\angle{BAC}) =\dfrac{4\sigma^2-\abs{\bm{X}}^2 -\abs{T}^2}{2\abs{\bm{X}}\abs{T}} $.
Choosing now the change of coordinates $\epsilon = \frac{\abs{T}^2+ 2\abs{T}\abs{\bm{X}}\cos\theta + \abs{\bm{X}}^2}{4\sigma^2}$ we obtain
\begin{align}
    &\dd\epsilon =  \frac{- 2\abs{T}\abs{\bm{X}}\sin\theta}{4\sigma^2}\dd\theta,\\
    &\theta_0  \rightarrow \epsilon_0 = 1,\\
    &\theta_1  \rightarrow \delta_{_-} =  \Big(\frac{\abs{\bm{X}}- \abs{T} }{2\sigma}\Big)^2.
\end{align}{}
We obtain then 
\begin{equation}
\begin{split}
    \xi_{ij} &=-  \tilde\lambda_i\tilde\lambda_j\frac{3\sqrt{2\pi}}{2\sigma\abs{\bm{X}}} \int_{1}^{\delta_{_-}} d\epsilon (1-\epsilon^{1/2})^2(2+\epsilon^{1/2}) = \frac{4\pi \sigma^2}{ \abs{\bm{X}}}(2\epsilon - 2\epsilon^{3/2} +\frac{2}{5}\epsilon^{5/2})    \Big|_{\delta_{_-}}^{1} \\
    & \implies \xi_{ij} = \lambda_i\lambda_j\frac{3\sqrt{2\pi}}{\sigma\abs{\bm{X}}} (\frac{1}{5}-\delta_{_-} + \delta_{_-}^{3/2} -\frac{1}{5}\delta_{_-}^{5/2}),
\end{split}   
\end{equation}
where we recall that $\delta_{_-} =\Big(\frac{\abs{\bm{X}}- \abs{T} }{2\sigma}\Big)^2$.

Now we consider the case when $\abs{T} < 2\sigma$.
When $2\sigma-\abs{T}\leq \abs{\bm{x}} \leq \abs{T}+2\sigma$, there is a partial intersection between \textit{S} and \textit{C}. 
Thus the calculations in the previous case hold exactly in the same fashion.
Now, when $\abs{\bm{x}} < 2\sigma-\abs{T} $, \textit{S} is completely engulfed by \textit{C}. 
For this case we need to evaluate the integral in \eqref{Yab integral}, within the limits 0 to $\pi$. 
Thus $\xi_{ij}$ for this case turns out to be
\begin{equation}
     \xi_{ij} = \tilde\lambda_i\tilde\lambda_j\frac{3\sqrt{2\pi}}{\sigma\abs{\bm{X}}} (\delta_{_+} - \delta_{_+}^{3/2} +\frac{1}{5}\delta_{_+}^{5/2} -\delta_{_-} + \delta_{_-}^{3/2} -\frac{1}{5}\delta_{_-}^{5/2}),
\end{equation}
where $\delta_{_-}= \Big(\frac{\abs{\bm{X}}- \abs{T} }{2\sigma}\Big)^2$ and $\delta_{_+}= \Big(\frac{\abs{\bm{X}} + \abs{T} }{2\sigma}\Big)^2$.

\subsubsection{Real part}
We now calculate $\zeta_{ij}$ which is given by
\begin{equation}
    \zeta_{ij}= \tilde\lambda_i\tilde\lambda_j 2\sqrt{\dfrac{2}{\pi}}\int\int \dd^3\bm{x}\;\dd^3\vb{z}\:\frac{F_j(\bm{z})F_i(\bm{x}+\bm{z})}{\abs{\bm{x}-\bm{X}}^2 - T^2}.
\end{equation}

Doing the $z$ integral just like when calculating the imaginary part, we obtain: 

\begin{equation}
    \zeta_{ij}= \tilde\lambda_i\tilde\lambda_j\frac{3}{\pi\sqrt{2\pi}\sigma^3} \int_{x\leq 2\sigma} \dd^3\bm{x} \Big(1-\frac{x}{2\sigma}\Big)^2\Big(2+\frac{x}{2\sigma}\Big)\frac{1}{\abs{\bm{x}-\bm{X}}^2 - T^2}.
\end{equation}

To calculate the above we use spherical coordinates. 
Without loss of generality, we can choose $\bm{X} = (0,0,\abs{\bm{X}})$.  We observe that the original volume of integration is a sphere with radius $2\sigma$ centered at the origin. We parametrize $\bm{x}$  as $\bm{r} = r(\cos\phi\sin\theta, \sin\phi\sin\theta, \cos\theta)$ with $\phi\in(0,2\pi)$ and $\theta\in(0, \pi)$. We observe that:
\begin{align}
    \abs{\bm{x}-\bm{X}}^2 = r^2- 2r\abs{\bm{X}}\cos\theta + \abs{\bm{X}}^2.
\end{align}{}

The integral, regarding the symmetry in the x-z axis,  thus becomes: 
\begin{equation}
\begin{split}
    \zeta_{ij} = \tilde\lambda_i\tilde\lambda_j\frac{3\sqrt{2\pi}}{\pi\sigma^3}  \int_0^{2\sigma} \dd r\int_{0}^{\pi} \dd\theta \sin(\theta) \Big(1-\frac{r}{2\sigma}\Big)^2\Big(2+\frac{r}{2\sigma}\Big)\frac{r^2}{ r^2- 2r\abs{\bm{X}}\cos\theta + \abs{\bm{X}}^2 - T^2}.
\end{split}   
\end{equation}

Choosing now the change of coordinates $\epsilon = -\cos\theta$ we obtain: 

\begin{equation}
\begin{split}
    \zeta_{ij} = \tilde\lambda_i\tilde\lambda_j\frac{3\sqrt{2\pi}}{\pi\sigma^3}  \int_0^{2\sigma} \dd r  \Big(1-\frac{r}{2\sigma}\Big)^2\Big(2+\frac{r}{2\sigma}\Big)r^2\int_{-1}^{1} \dd\epsilon \frac{1}{r^2 +2r\abs{\bm{X}}\epsilon + \abs{\bm{X}}^2 - T^2}.
\end{split}   
\end{equation}

The integral in $\epsilon$ has the form $\frac{1}{a\epsilon + b}$, with $a\coloneqq 2 r \abs{\bm{X}}$ and $b\coloneqq r^2 + \abs{\bm{X}}^2 - T^2$. This function has a pole at $\epsilon_{\textrm{p}} = - \frac{b}{a}$. However, its also odd about $\epsilon_{\textrm{p}}$, and thus we can get rid of this singularity.
The location of the pole
depends on the relation between $2\sigma, \abs{\bm{X}}$ and $\abs{T}$, which leads to three mutually exclusive cases as shown below. To make the results easy to follow, we use the notation $g(r) = r^2\Big(1-\frac{r}{2\sigma}\Big)^2\Big(2+\frac{r}{2\sigma}\Big)$ and $K = \tilde\lambda_i\tilde\lambda_j\frac{3\sqrt{2\pi}}{\pi\sigma^3} $. 

\begin{itemize}
\item \textbf{Case 1:} $2\sigma< \abs{\abs{\bm{X}}-\abs{T}}$.
In this case $\epsilon_{\textrm{p}}< -1$ for $r\in (0,2\sigma)$. Therefore we have 

\begin{align}
     \zeta_{ij} = K \int_0^{2\sigma} \dd r  g(r)\int_{-1}^{1} \dd\epsilon \dfrac{1}{a\epsilon + b}.
\end{align}

\item \textbf{Case 2:} $\abs{\abs{\bm{X}}-\abs{T}}<2\sigma< \abs{\bm{X}}+\abs{T}$.
In this case $\epsilon_{\textrm{p}}< -1$ for $r\in (0,\abs{\abs{\bm{X}}-\abs{T}})$ and  $-1<\epsilon_{\textrm{p}}<0$ for $r\in (\abs{\abs{\bm{X}}-\abs{T}}, 2\sigma)$. Therefore we have 

\begin{equation}
\begin{split}
     \zeta_{ij} = K \Bigg[\int_0^{\abs{\abs{\bm{X}}-\abs{T}}} \dd r  g(r)\int_{-1}^{1} \dd\epsilon \dfrac{1}{a\epsilon + b} + \int_{\abs{\abs{\bm{X}}-\abs{T}}}^{2\sigma} \dd r  g(r) \int_{2\epsilon_{\textrm{p}}+1}^{1} \dd\epsilon \dfrac{1}{a\epsilon + b} \Bigg].
\end{split}   
\end{equation}

\item \textbf{Case 3:}  $\abs{\bm{X}}+\abs{T}<2\sigma$.
In his case $\epsilon_{\textrm{p}}< -1$ for $r\in (0,\abs{\abs{\bm{X}}-\abs{T}})$,   $-1<\epsilon_{\textrm{p}}<0$ for $r\in (\abs{\abs{\bm{X}}-\abs{T}}, \abs{\bm{X}}+\abs{T})$ and  $\epsilon_{\textrm{p}}< -1$ for $r\in (\abs{\bm{X}}+\abs{T}, 2\sigma)$. Therefore we have 

\begin{equation}
\begin{split}
     \zeta_{ij} = K \Bigg[\int_0^{\abs{\abs{\bm{X}}-\abs{T}}} \dd r  g(r)\int_{-1}^{1} \dd\epsilon \dfrac{1}{a\epsilon + b} +  \int_{\abs{\abs{\bm{X}}-\abs{T}}}^{\abs{\bm{X}}+\abs{T}} \dd r  g(r) & \int_{2\epsilon_{\textrm{p}}+1}^{1} \dd\epsilon \dfrac{1}{a\epsilon + b} \\
    & +\int_{\abs{\bm{X}}+\abs{T}}^{2\sigma} \dd r g(r) \int_{-1}^{1} \dd\epsilon \dfrac{1}{a\epsilon + b} \Bigg].
\end{split}   
\end{equation}

\end{itemize}{}

The expression for $\zeta_{ij}$ is greatly simplified if $a=b$, in which case, $\abs{\bm{X}} = 0= T$ and we obtain: 
\begin{equation}
    \zeta_{ii}= \tilde\lambda_i^2\frac{3}{\pi\sqrt{2\pi}\sigma^3} \int_{x\leq 2\sigma} \dd^3\bm{x} \Big(1-\frac{x}{2\sigma}\Big)^2\Big(2+\frac{x}{2\sigma}\Big)\frac{1}{x^2} = \tilde\lambda_i^2\frac{12}{\sqrt{2\pi}\sigma^3} \int_{0}^{2\sigma} \dd x \Big(1-\frac{x}{2\sigma}\Big)^2\Big(2+\frac{x}{2\sigma}\Big).
\end{equation}

\subsection{Gaussian smearing function}
Now we take the smearing function to be a normalised gaussian function in 3 dimensions:
\begin{equation}
    F(\bm{x}) = \frac{1}{(2\pi\sigma^2)^{3/2}}\exp[-\frac{\bm{x}^2}{2\sigma^2}].
\end{equation}
We wish to evaluate the imaginary part of the equation \eqref{Tab1} to obtain $\xi_{ij}$,using the above smearing function. 
\begin{equation}
    \xi_{ij}= \tilde\lambda_i\tilde\lambda_j\sqrt{2\pi}\int\int \dd^3\bm{x}\;d^3\vb{z}\:F_j(\bm{z})F_i(\bm{x}+\bm{z})\frac{T\delta(\abs{\bm{x}-\bm{X}}-\abs{T})}{\abs{T}\abs{\bm{x}-\bm{X}}}.
\end{equation}
While doing so, we first solve the integral over $\vb{z'}$ as we had done in the case of a hard-sphere smearing function.
This integral is a Gaussian integral and is readily performed.
Thus we obtain,
\begin{equation}
\label{xiijImaginary}
     \xi_{ij}= \frac{\tilde\lambda_i\tilde\lambda_j\sqrt{2\pi}}{(4\pi\sigma^2)^{3/2}}\int \dd^3\bm{x}\: \exp[-\frac{\abs{\bm{x}}^2}{4\sigma^2}] \frac{T\delta(\abs{\bm{x}-\bm{X}}-\abs{T})}{\abs{T}\abs{\bm{x}-\bm{X}}}.
\end{equation}
Performing a change of variable to $\vb{r}=\bm{x}-\bm{X}$, we get
\begin{equation}
     \xi_{ij}= \frac{\tilde\lambda_i\tilde\lambda_j\sqrt{2\pi}}{(4\pi\sigma^2)^{3/2}}\int \dd^3\vb{r}\: \exp[-\frac{\abs{\vb{r+X}}^2}{4\sigma^2}] \frac{T\delta(\abs{\vb{r}}-\abs{T})}{\abs{T}\abs{\vb{r}}}.
\end{equation}
Without loss of generality, we can choose $\bm{X} = (0,0,\abs{\bm{X}})$. 
To evaluate the integral we use spherical coordinates, and parametrize $\bm{r}$  as $\bm{r} = r(\cos\phi\sin\theta, \sin\phi\sin\theta, \cos\theta)$ with $\phi\in(0,2\pi)$ and $\theta\in(0, \pi)$.
Recall that:
\begin{align}
    \abs{\bm{r}+\bm{X}}^2 = r^2+ 2r\abs{\bm{X}}\cos\theta + \abs{\bm{X}}^2.
\end{align}
Since we have the term $\delta(r-\abs{T})$, the volume of integration parametrized by $r, \phi, \theta$ will become a spherical shell centered at $\bm{0} = (0,0,0)$ parametrized by $\phi, \theta$.
Moreover the intergrand depends only on $\theta$ and $r$, so the integral over $\phi$ just gives a constant of $2\pi$. Subsequently we obtain,
\begin{equation}
    \xi_{ij}= \frac{\tilde\lambda_i\tilde\lambda_jT}{2\sqrt{2}\sigma^3}\exp[-\frac{T^2+\abs{\bm{X}}^2}{4\sigma^2}]\int_{-1}^1 \dd\cos{\theta} \;\exp[-\frac{\abs{T}\abs{\bm{X}}\cos{\theta}}{2\sigma^2}].
\end{equation}
Now we have an elementary integration of the exponential function. 
The final expression for $\xi_{ij}$ for the gaussian smearing function is
\begin{equation}
    \xi_{ij}= \frac{\tilde\lambda_i\tilde\lambda_jT}{\sqrt{2}\sigma\abs{\bm{X}}\abs{T}}\Bigg[ \exp[-\frac{(\abs{T}-\abs{\bm{X}})^2}{4\sigma^2}] - \exp[-\frac{(\abs{T}+\abs{\bm{X}})^2}{4\sigma^2}] \Bigg].
\end{equation}
To calculate the real part $\zeta_{ij}$ which is given by
\begin{equation}
    \zeta_{ij}= \tilde\lambda_i\tilde\lambda_j 2\sqrt{\dfrac{2}{\pi}}\int\int \dd^3\bm{x}\;\dd^3\vb{z}\:\frac{F_j(\bm{z})F_i(\bm{x}+\bm{z})}{\abs{\bm{x}-\bm{X}}^2 - T^2},
\end{equation}
We perform the integral over $z$ in the same way as in \eqref{xiijImaginary} while calculating the imaginary part $\xi_{ij}$. We obtain,
\begin{equation}
     \zeta_{ij}= \frac{\tilde\lambda_i\tilde\lambda_j2\sqrt{2}}{\sqrt{\pi}(4\pi\sigma^2)^{3/2}}\int \dd^3\bm{x}\: \exp[-\frac{\bm{x}^2}{4\sigma^2}] \frac{1}{\abs{\bm{x}-\bm{X}}^2 - T^2}.
\end{equation}
Subsequently we numerically integrate the above expression for particular choice of parameters to obtain all the results concerning Gaussian smearing functions.

\section{An illustrative example of entanglement structures with four qubits}
\label{Appendix:Example}
In our study involving multiple qubits and a field, we encounter this peculiar scenario, where
one of the detectors (Bob) ends up in a maximally mixed state, without having any bipartite entanglement with other detectors or maximal entanglement with the field. 
This raises the suspicion that there might be genuine multipartite entanglement among the detectors and the field. 
To understand how this might happen in a simpler scenario, here we present a toy model with a four qubit state $\ket{\Psi_{\textsc{ABIF}}}$ for Alice, Bob, interloper and Field (where a qubit represents the field), to mimic our correlation sabotaging setup we impose the following  conditions:
\begin{enumerate}
    \item $\ket{\Psi_{\textsc{abif}}}$ is a pure state
    \item The joint state of Alice and Bob is a product state of the form $\rhoh_{\textsc{a}}\otimes\frac{\openone_{\textsc{b}}}{2}$.
    \item interloper and Alice may have correlations between them as do interloper and Bob, but they are not entangled.
\end{enumerate}

We now consider the state, 
\begin{align}
    \ket{\Psi_\textsc{aibf}} = \frac{1}{\sqrt{2}}(a\ket{0_\ta}+c\ket{1_\ta})\ket{0_\ti}\ket{\Phi^{+}_{\tb\tf}}+\frac{1}{\sqrt{2}}(b\ket{0_\ta}+d\ket{1_\ta})\ket{1_\ti}\ket{\Psi^{+}_{\tb\tf}},
\end{align}
where $\ket{\Phi^{+}_{\tb\tf}}=\frac{1}{\sqrt{2}}(\ket{00}+\ket{11})$ and $\ket{\Psi^{+}_{\tb\tf}}=\frac{1}{\sqrt{2}}(\ket{01}+\ket{10})$ are Bell states.
The coefficients $a,b,c,d$ are complex numbers that satisfy $|a|^2+|b|^2=1$ and $|c|^2+|d|^2=1$. 
We can now proceed to calculate various bipartite partial states and check the validity of the listed conditions. 
Firstly we calculate $\rhoh_{\tab}$ and see that 
\begin{align}
    \rhoh_{\tab} = \frac{1}{2}\left(\begin{array}{cc} 1 & ac^*+bd^* \\ a^*c+b^*d & 1\end{array} \right) \otimes \frac{\openone_{\tb}}{2},
\end{align}
validating condtion 2.
 
We calculate the bipartite partitions containing the interloper to show condition 3 holds as well. 
We find that
\begin{align}
    \rhoh_{\ta\ti}= \frac{1}{2}(a\ket{0_\ta}+b\ket{1_\ta})(a^*\bra{0_\ta}+b^*\bra{1_\ta})\otimes\ket{0_\ti}\bra{0_\ti} + \frac{1}{2}(c\ket{0_\ta}+d\ket{1_\ta})(c^*\bra{0_\ta}+d^*\bra{1_\ta})\otimes\ket{1_\ti}\bra{1_\ti},
\end{align}
which is a separable state. Hence there is no bipartite entanglement between Alice and the interloper. 
We shall also see that Bob and interlopers state is also separable.
By a straightforward tracing over Alice's and Field's
qubits we find:
\begin{align}
    \rhoh_{\ti\tb}= \frac{1}{4}
    \left(
      \begin{array}{cccc}
        |a|^2+|c|^2 & 0 & 0 & ab^*+cd^*  \\
        0 & |a|^2+|c|^2 & ab^*+cd^* & 0 \\
        0 & a^*b+c^*d & |b|^2+|d|^2 & 0 \\
        a^*b+c^*d  & 0 & 0 & |b|^2+|d|^2
      \end{array}
    \right)
\end{align}
We notice that this state is unchanged when it is partially transposed w.r.t Bob's subsystem. 
However as the density matrix itself must have non-negative eigenvalues, we conclude that the negativity $\mathcal{N}(\rhoh_{\ti\tb})=0$, which for a system of two qubits is a sufficient proof of separability.
Thus we have established that the interloper doesn't have bipartite entanglement with the  condition 3 holds true as well for our example. 

We can further try to understand how the `field qubit' is entangled with the detector qubits. In our example the state $\ket{\Psi_{\ta\ti\tb\tf}}$ is symmetric under the exchange of Bob and field's qubit. Hence the states $\rhoh_{\ta\tf}, \rhoh_{\ti\tf}$ are the same as $\rhoh_{\ta\tb}, \rhoh_{\ti\tb}$ respectively, showing that Alice and interloper have no bipartite entanglement with the field qubit.
However that is not the case for the Bob-Field partition. We find Bob's and Field's partial state to be:
\begin{align}
    \rhoh_{\ti\tb}= \frac{1}{4}
    \left(
      \begin{array}{cccc}
        |a|^2+|c|^2 & 0 & 0 & |a|^2+|c|^2  \\
        0 & |b|^2+|d|^2 & |b|^2+|d|^2 & 0 \\
        0 & |b|^2+|d|^2 & |b|^2+|d|^2 & 0 \\
        |a|^2+|c|^2  & 0 & 0 & |a|^2+|c|^2 
      \end{array}
    \right)
\end{align}
The negativity of this state turns out to be 
\begin{align}
    \mathcal{N}(\rhoh_{\tb\tf}) &= \frac{1}{4}\big||b|^2+|d|^2-|a|^2-|c|^2\big| \\
    & = \frac{1}{4}\big||b|^2+|d|^2+|a|^2+|c|^2 -  2(|a|^2+|c|^2)\big| \\
    & \frac{1}{4}\big|2-2(|a|^2+|c|^2))\big|\leq 1/2
\end{align}
Since the maximum value of negativity for 2 qubits is $1/2$, in general Bob and Field have some non zero bipartite entanglement between them, but are not maximally entangled.
This demonstrates that in our toy example satisfying all the essential features of our actual detector-field state, how it is possible for Bob to be in a maximally mixed state without being maximally entangled with any single subsystem.

\end{widetext}

\bibliography{references}
\end{document}